
ocumentstyle[aps,preprint,eqsecnum]{revtex}
\def\gsim
 {\raise2pt\hbox
 {$\displaystyle{}\mathrel{\mathop>_{\raise1pt\hbox{$\sim$}}}{}$}}
\def\lsim
 {\raise2pt\hbox
{$\displaystyle{}\mathrel{\mathop<_{\raise1pt\hbox{$\sim$}}}{}$}}
\begin{document\title{Macroscopic Quantum Tunneling \\ of a Domain Wall \\
in a Ferromagnetic Metal}
\author{Gen Tatara and Hidetoshi Fukuyama }
\address{        Department of Physics, University of Tokyo,
        7-3-1 Hongo, Tokyo 113, Japan }
\date{\today}
\maketitle
\newcommand{\beq}{\begin{equation}}
\newcommand{\eeq}{\end{equation}}
\newcommand{\beqa}{\begin{eqnarray}}
\newcommand{\eeqa}{\end{eqnarray}}

%
\newcommand{\sigbf}{\mbox{\boldmath$\sigma$}}
\newcommand{\Mv}{{\bf M}}
\newcommand{\nv}{{\bf n}}
\newcommand{\Jv}{{\bf J}}
\newcommand{\kv}{{\bf k}}
\newcommand{\qv}{{\bf q}}
\newcommand{\xv}{{\bf x}}
\newcommand{\half}{\frac{1}{2}}
\newcommand{\kf}{k_{\rm F}}
\newcommand{\kfu}{k_{{\rm F}\uparrow}}
\newcommand{\kfd}{k_{{\rm F}\downarrow}}
\newcommand{\kfs}{k_{{\rm F}\sigma}}
\newcommand{\ef}{\epsilon_{\rm F}}

\newcommand{\efs}{\epsilon_{{\rm F}\sigma}}
\newcommand{\eks}{\epsilon_{{\bf k}\sigma}}
\newcommand{\ekqs}{\epsilon_{\kv+\qv,\sigma}}
\newcommand{\ekd}{\epsilon_{\kv\downarrow}}
\newcommand{\ekqu}{\epsilon_{\kv+\qv,\uparrow}}
\newcommand{\fks}{f_{\kv\sigma}}
\newcommand{\fkqs}{f_{\kv+\qv,\sigma}}

\newcommand{\fkqu}{f_{\kv+\qv,\uparrow}}
\newcommand{\cel}{c_{\xv\sigma}}
\newcommand{\ael}{a_{\xv\sigma}}
\newcommand{\cd}{c_{\xv\sigma}^\dagger}

\newcommand{\thx}{\theta_{\xv}}
\newcommand{\phx}{\phi_{\xv}}
\newcommand{\Hu}{H_{\rm U}}

\newcommand{\Huf}{H_{\rm U}^{(\rm fast)}}
\newcommand{\Ms}{M_0}
\newcommand{\Cpm}{C_{+-}}

\newcommand{\oml}{\omega_\ell}

\newcommand{\DSone}{\Delta S^{(1)}}
\newcommand{\DStwo}{\Delta S^{(2)}}
\newcommand{\nzero}{n_0}
\newcommand{\Awall}{A_{\rm w}}
\newcommand{\Spin}{S}
\newcommand{til}{\tilde k_0}
\newcommand{\lamtil}{\tilde \lambda}
\newcommand{\mels}{m'}
\newcommand{\nels}{n'}
\begin{abstract}
The macroscopic quantum tunneling of a planar domain wall in a ferromagnetic
metal is studied by use of an instanton method.
Based on the Hubbard model, the effective action of the magnetization is
derived within the assumption of slow dependences on space and time.
The resulting action is formally similar to that of
a ferromagnetic Heisenberg model but with a term non-local in time that
describes the dissipation due to the itinerant electron.
The crucial difference from the case of the insulator is the presence of the
ohmic dissipation even at zero temperature due to the gapless Stoner
excitation.
The reduction of the tunneling rate due to the dissipation is calculated.
The  dissipative effect is found to be very large for a thin domain wall
with thickness of a few times the lattice spacing, but is negligible for
a thick domain wall.
The results are discussed in the light of recent experiments on
ferromagnets with strong anisotropy.


\end{abstract}
\newpage
\section{Introduction}
\label{SECint}
The quantum phenomena occuring in a macroscopic scale, {\it i.e.,}
macroscopiquantum phenomena (MQP), are
intensively investigated both theoretically and experimentally for about
a decade after the seminal paper by Caldeira and Leggett\cite{CL}.
In macroscopic systems, the coupling to the environment, and hence
dissipation, may become significant.
Caldeira and Leggett presented a formalism that can incorporate
the dissipation by use of the imaginary time path integral.
There the dissipation is expressed by the action non-local in time.
Based on this formulation, Caldeira and Leggett investigated the quantum
tunneling of a macroscopic variable (MQT), and found that the dissipation
generally reduces the tunneling rate.
Their prediction turned out to be in excellent agreement with the experiment
carried out on Josephson junction\cite{VW}.
Another important subject in the macroscopic quantum phenomena is the
macroscopic quantum coherence (MQC), in which the potential for the
macroscopic variable has several equivallent minima and the coherence among
minima arises from the quantum tunneling\cite{BM,LCD}.
A typical example of MQC is the diffusion process of a heavy particle in
metal\cite{Kondo,ATF}, where the dissipation due to conduction electron
plays crucial roles.

Besides these MQP that have been explored extensively so far,
there has been growing interest in MQP in magnetic
systems\cite{Gun,SCB,Bar,TZB,LDG2} in recent years, owing mainly to the
development of technology in mesoscopic physics in nanoscale structures.
In magnetic systems, the processes due to MQP are the following.
In the case of a small ferromagnet with an easy axis, the magnetization is
uniform ({\it i.e.,} single domain).
The direction of the magnetization vector
may be reversed by applying a magnetic field in the opposite direction to
the magnetization.
At low temperature, this process is due to the quantum tunneling of a
macroscopic number of spins through the anisotropy energy
barrier.\cite{BL,ES,HS,CG}
Also MQC of the magnetization vector or the N\'eel vector may occur in both
ferromagnetic\cite{CG} and antiferromagnetic\cite{BC,KZ} single domain
magnets.
A larger ferromagnet contains several or more domains to reduce the
magnetostatic energy due to  magnetization.
In this case, the nucleation process of domains\cite{CGnuc} at low
temperatures
in the presence of a magnetic field  can be considered as MQT.
Domain walls are generally pinned by impurities, and by applying the
magnetic
field, the wall moves by tunneling through the energy barrier produced
by pinning centers.
This is the depinning of a domain wall via MQT\cite{Sta,CIS}.

The tunneling in single domain ferromagnet have been theoretically
investigated more than
thirty years ago by Bean and Livingston\cite{BL}. This problem was
investigated in more detail by Enz and Schilling\cite{ES} and by Hemmen and
S\"{u}t\H{o}\cite{HS}.
A transparent description of this MQT was given and the possibility of
experimental observation was discussed by Chudnovsky and Gunther\cite{CG} in
1988. By use of the instanton method they calculated the tunneling rate of
the magnization vector in a
small single domain particle assuming several forms of anisotropy energy,
and found that it is large enough to be obeserved in experiments
in a particle of a size 100\AA\ at $T\lsim0.06K$.
To see if this MQT  is  really observable, however, the dissipative effect
arising from the coupling to the environment must be estimated as pointed
out by Caldeira and Leggett\cite{CL}.
As the most obvious source of dissipation, the magnetoelastic coupling to
phonons was considered, but the effect turned out be negligible in actual
situations.\cite{CG,GK,Sim}
The effects of spin wave have not been discussed, but are expected to be
negligible since this degree of freedom
is frozen out at very low temperature due to the anisotropy gap.
Recently it was found that the dissipation due to hyperfine coupling to the
nuclearspin is significant.\cite{Gargnuc}
Rotatin of N\'{e}el vector in single domain antiferromagnet has also been
discussed based on the anisotropic $\sigma$ model\cite{BC,KZ}. This MQC would
be more favorable for observation than that in ferromagnets because of larger
quantum fluctuation in antiferromagnets\cite{BC}. A brief discussion on the
effect of dissipation was given\cite{BC}, but on only phenomenological ground.
Chudnovsky and Gunther\cite{CGnuc} also indicated that the qunatum nucleation
of domains would be observable for highly anisotropic ferromagnetic materials
by applying a magnetic field.
As regards the tunneling of the spin vector, it has been demonstrated that
the topological term in the action can lead to destructive interference among
different paths\cite{LDG,DH}, and there are several studies which focused on
this spin-parity effect.\cite{LDG2,Gtop,DH2,CD}
This effect is related to the Kramers degeneracy of the ground state of
half-odd-integer spin in the absence of magnetic field.

The MQT of a ferromagnetic domain wall was investigated by Stamp\cite{Sta}.
His consideration on an insulating magnet in three dimensions is based on the
ferragnetic Heisenberg model with a uniaxial anisotropy.
By use of a classical solution of planar domain wall, the
hamiltonian describing the position of the  wall is obtained.
The pinning potential by a small defect is assumed to be of a short range.
In the presence of a magnetic field $H$, the position of the wall at the
pinning center becomes metastable.
The barrier height is reduced as the field is increased and finally
vanishes at the coercive field, $H_c$.
Thexperiments of the tunneling of the wall from the metastable minima are
carried out in a magnetic field close to this coercive field for the rate
to be large enough to be observed.
Stamp obtained the expression of the tunneling rate
as functions of macroscopic variables of the magnet; {\it e.g.,} the coercive
field, saturation magnetization, the number of spins in the wall and the
external field, and concluded that
depinning of a wall due to MQT is observable  even for a large wall
containing about $10^{10}$ spins, if dissipation can be neglected.

As a source of dissipation in this MQT of domain walls, he considered the
magnon (spin wave) and phonon.
These degrees of freedom do not couple linearly to the wall coordinate,
unlike in the model of Caldeira and Leggett.
The effects of magnon are calculated by the Holstein-Primakoff transformation
around the domain wall background.
The dissipative effect is calculated from the imaginary part of the
self-energy of two and three magnon processes.
Because of the magnon gap due to the anisotropy, the ohmic dissipation
vanishes at absolute zero  and is quantitatively negligible at low
temperature.
The phonon was also shown to be irrelevant.
The effects of conduction electrons were briefly touched by Chudnovsky
it et al.}\cite{CIS} on a phenomenological ground.

In order to observe these MQP in magnetic systems, the temperature must
be sufficiently low for the thermal activation process to be suppressed, and
at the same time, the tunneling rate needs to be large enough compared to
the time of observation.
Because of these requirements, the MQP in magnetic systems have not long been
observed in conventional bulk magnetic metals.
The theoretical studies\cite{CG,CGnuc} indicates that highly anisotropic
materials (such as those containing rare-earth) are suitable for observing
the MQP, since in these cases, the crossover temperature, $T_{\rm co}$ from
the thermal to the quantum regime can be higher.
Recently random magnets has been discussed to be suitable for MQT, since the
condition of low energy barrier is easier to be realized in these systems
\cite{Chud,TZC}
Experimental indication of MQP in magnetic system\cite{Bar,TZB} was first
obtained for a bulk single crystal of SmCo$_{3.5}$Cu$_{1.5}$\cite{UB}.
This material has a large uniaxial anisotropy and the width of domain wall is
about $12$\AA.
It was found in the measurement of the magnetization relaxation that the
relaxation rate goes to a constant value below $T\lsim10$K.
This tendency suggests the motion of a small portion of domain wall via MQT.
Later, the magnetization relaxation was measured on a small ferromagnetic
particles of Tb$_{0.5}$Ce$_{0.5}$Fe$_2$ of diameter of 150\AA.\cite{PSB}
The characteristic feature of this experiment is that the particle of this
size is likely to contain only one domain wall.
The temperature independent relaxation was observed for $T\lsim0.6$K, and
the result was claimed to be consistent with theory\cite{Sta} of MQT of
a domain wall in a ferromagnet without dissipation.
The tunneling of magnetization vector of a single domain particle is claimed
to have been observed in ferromagnetic FeC particles of 40\AA \ in
diameter\cite{BTL} below 1K. The results were consistent with theoretical
predictions\cite{CG}.
As an attempto observe MQC in single domain magnets, the measurement
of AC susceptibility at low temperature was carried out on nanometer-scale
ferromagnets of Fe(CO)$_5$, which have been fabricated by use of a scaning
tunneling
microscope.\cite{AMG} A resonance was seen at $\omega\sim 400$Hz, suggesting
the oscillation of the magnetization vector between two easy directions via
MQC. However, because of the quantitative discrepancy between the
experimental
result and the
theoretical prediction\cite{CG,Garg}, MQC explanation of the resonance is
suspicious\cite{AMG,SCB,Garg}.
It is also reported in a similar experiment that MQC of a N\'eel vector was
observed for single domain antiferromagnetic particles of horse spleen
ferritin\cite{ASG}. It should be pointed out, however, that the MQC
interpretation of the result is also still controversial\cite{Garg,ASG2}.
Mesurements of the magnetic relaxation have been carried out on a multilayers
 of SmCo\cite{ZBR} and SmFe\cite{ZBR2}, whose results indicate the existence
of the nucltion process via MQT below about 3K.
Recently, more datailed studies of domain wall MQT has been carried out
in so called the domain wall junctions\cite{Bar}, where the pinning
potential for the domain wall can be controled by
choosing the material between two layers.
Magnetic relaxation due to MQT has also been indicated in random
magnets\cite{TZC}.

Although the materials employed in these experiments are generally matallic,
there has been
no theoretical work paying full attention to this fact.
The aim of this paper is  therefore to investigate the MQT in a
ferromagnetic metal based on a itinerant electron model, {\it i.e.,} the
Hubbard model. We consider the case of absolute zero, $T=0$ ,
since we are interested only in the quantum tunneling present at low
temperature.
The calculation is carried out in the continum.
The resul would be
valid even for a thin domain wall with width of a few times lattice
constant, at least by order of magnitude.
The magnetization vector is expressed as an expectation value
of the electron spin operator.
For the treatment of a slowly varying field like a domain wall, the locally
 rotated frame for the electron is convenient.
By use of this frame, the effective action that describes
the low energy dynamics of magnetization vector is obtained
by integrating out the electron degrees of freedom
in the imaginary time path integral.
As far as low energy behavior is concerned,
the moation of the magnitude of the magnetization is irrelevant, and
only its direction can be a dynamical variable.
Then the part of the effective action that is local in time ({\it i.e.,}
instantaneous) has the same form as that of  a ferromagnetic Heisenberg
model; {\it i.e.,} there is no formal difference from the case of the
insulator but the parameters have different physical origin.
On the other hand, the non-local ({\it i.e.,} retarded) part of the
effective action, which describes the dissipative effect of itinerant
electron on the motion
of the magnetization, is  crucially different from the case of an insulator.
Due to the Stoner excitaion, which is the gapless excitation of spin flip
across the fermi surface, the ohmic dissipation is present even at zero
temperature.
It ishown that this dissipation reduces the tunneling rate very much
particularly for a thin
domain wall of thickness comparable to the inverse of the difference
of the fermi momenta $(\kfu-\kfd)^{-1}$.
In the case of strong ferromagnet, on the other hand, dissipation resulting
from the Stoner excitation is reduced. In such a case the existence of
non-magnetic band in actual systems is to be taken into account.
In contrast to these effects from the Stoner excitation, the effects of ohmic
dissipation due to charge current (eddy current) are found to be negligible
in a meso- or microscopic wall we are interested in.

This paper is organized as follows.
\S\ref{SECeff} is devoted to the derivation of the effective action for
the magnetization vector on the basis of the Hubbard model.
In \S\ref{SECdw} the MQT of a domain wall is studied based on this action.
The expression of the tunneling rate including the
dissipative effect from itinerant electron is obtained there.
The dissipative effects resulting from the non-magnetic band are calculated
in
\S\ref{SECsd}, and in \S\ref{SECeddy} the  effects due to the eddy current
are estimated.
Discussions and conclusion are given in \S\ref{SECdisc} and \S\ref{SECconc}.

This paper is a detailed and extended version of the preceeding
letter\cite{TF}.
\section{Derivation of an Effective Action}
\label{SECeff}
\subsection{Model}
We consider the Hubbard model whose Lagrangean
in the imagnary time path integral is given by
\begin{equation}
L=\int d^3 x \sum_{\sigma} \left( \cd (\partial_\tau-\epsilon_{\rm F})
\cel+\frac{1}{2m} |\nabla \cel|^2 \right)
   +\Hu , 		\label{L0}
\end{equation}
where $\cel$ is an electron operator at site $\xv$ with spin $\sigma(=\pm)$
and $n_{\xv\sigma}\equiv\cd\cel$.
The term $\Hu$ represents the on-site Coulomb repulsion
\beq
\Hu\equiv U \int d^3 x  n_{\xv\uparrow}n_{\xv\downarrow}  . \label{HU0}
\eeq
The chemical potential is denoted by $\epsilon_{\rm F}$.
For simplicity, a parabolic dispersion with mass $m$ is assumed for band
energy.
The interaction $\Hu$ can be expressed in terms of the electron spin
oerator using the identity:
beqa
n_{\xv \uparrow}n_{\xv \downarrow} &=& -\half (n_{\xv \uparrow}-n_{\xv
\downarrow})^2
   +\half(n_{\xv \uparrow}^2+n_{\xv \downarrow}^2) \nonumber\\
 &=& -\half (c^\dagger \sigma_z c)_\xv^2 +\half\sum_\sigma n_{\xv\sigma},
\label{Unn}
\eeqa
where  we have suppressed the spin indexes
($(c^{\dagger}  \bbox{\sigma}  c)_\xv
 \equiv \sum_{\sigma\sigma'} \cd \sigbf_{\sigma\sigma'} \cel$)
and ${\bf \sigbf} =\sigma_i (i=x,y,z) $ are the Pauli matrixes.
The last term is absorbed in the definition of the chemical potential
$\epsilon_{\rm F}$In the momentum space, $\Hu$ is then written as
\beq
\Hu=-\frac{U}{2V}\sum_{\qv\kv\kv'}
(c^\dagger_{\kv+\qv} \sigma_z c_\kv )(c^\dagger_{\kv'-\qv}
\sigma_z c_{\kv'} ),
\eeq
where $V$ is the volume of the system and
$\sum_{\kv}\equiv V\int d^3\kv/(2\pi)^3$.
Due to the rotational invariance, we can write $\Hu$  using
 a (slowy varying) unit vector $\nv(\xv,\tau)$ ($|\nv|^2=1$) pointing to
some arbitrary direction as (similarly to eq.(\ref{Unn}))
\beq
\Hu= -\frac{U}{2}\int d^3 x [(c^\dagger {\bf \sigbf} c)_\xv\cdot \nv(\xv)
]^2
.  \label{HU1}
\
The vector $\nv(\xv)$ will represent the direction of the magnetization
vector.
For the rotational invariance to be preserved, the integration over
$\nv(\xv)$
 is needed in the path integral.
Introducing the magnitude of the magnetization
\beq
M(\xv)\equiv <(c^\dagger {\bf \sigbf} c)_\xv > \cdot \nv(\xv),  \label{Mdef}
\eeq
we write the Lagrangean as follows by use of the Hubbard-Stratonovich
transformation
\beq
L=\sum_{\kv\sigma} c^\dagger_{\kv \sigma}(\partial_\tau + \epsilon_\kv)
c_{\kv\sigma}
  -U\int d^3x\Mv(\xv) (c^\dagger {\bf \sigbf} c)_\xv
 +\frac{U}{2} \int d^3 xM(\xv)^2,
  \label{L1}
\eeq
where the magnetization vector is $\Mv(\xv)\equiv M(\xv) \nv(\xv)$.
The partition function is now written as
\beq
Z=\int{\cal D}c^\dagger{\cal D}c{\cal D}M{\cal D}\nv \delta(\nv^2-1)
  \exp\left(-\int_0^\beta d\tau L \right)  . \label{Z}
\eeq

The spatial variation of $\nv(\xv)$ accompanying with a
domain wall is much slower compared to
the inverse fermi momentum of the electron $\kf^{-1}$.
For the analysis of such a slowly varying field,
the local frame of electron, which was originally introduced by Korenman,
Murray and Prange\cite{Pra} in their "local-band description" of the
itinerant
ferromagnetism, is convenient. In this frame the $z$-axis
of the electron is chosen in the direction of the local magnetization vector
$\nv(\xv)$ (see Appendix \ref{AProt} for details on the locally rotated
frame).
The electron operator in the new frame $\ael$ is related to $\cel$ in the
origial one as
\beqa
 \left( \begin{array}{c} a_{\xv\uparrow} \\ a_{\xv\downarrow} \end{array}
 \right)
 = \left(\begin{array}{cc} \cos{\frac{\thx}{2}}  &
                           e^{-i\phx}\sin{\frac{\thx}{2}} \\
      e^{i\phx}\sin{\frac{\thx}{2}}   &  -\cos{\frac{\thx}{2}}  \end{array}
     \right)
    \left( \begin{array}{c} c_{\xv\uparrow} \\ c_{\xv\downarrow} \end{array}
  \right),
\label{loc}
\eeqa
where $(\thx(\tau),\phx(\tau))$ are the polar coordinates of vector
$\nv(\xv,\tau)$ (Fig. \ref{FIGpolar}).
In terms of $\ael$, the second term in eq.(\ref{L1}) is expressed as
\beq
-U\int d^3 xM(\xv)\nv(\xv) (c^\dagger {\bf \sigbf} c)_\xv
 =-U\int d^3 x M(\xv)(a^\dagger { \sigma_z} a)_\xv. \label{int}
\eeq
As a price of this transformation, the kinetic energy term
$c^\dagger \dot c + |\nabla c|^2/(2m)$ induces additional terms that describe
the interaction of electrons with space-time variation of the magnetization
vector. We neglect in the following calculation the spatial variation of the
magnitude $M(\xv)$ of the magnetization ({\it i.e.,} $M(\xv)\equiv M$), since
the fluctuation of $M(\xv)$ has a finite mass as will be discussed later.
Thus the relevant degree of freedom at low energy is only
the variation of the angle, ($\theta,\phi$), of the magnetization.
The interaction terms are
\beqa
\lefteqn{ H_{\rm int}=\int d^3 x \left[ \frac{i}{2}\dot{\phx}(1-\cos\thx)
 (a^\dagger\sigma_z a)_\xv
 \right. +\sum_{\pm} \half(\pm\dot{\thx}-i\sin\thx\dot{\phx})e^{\mp i\phx}
(a^\dagger\sigma_\pm a)_\xv
 }   \nonumber\\
&& +\frac{1}{4m}\left(\half (\nabla\thx)^2+(1-\cos\thx)(\nabla\phx)^2 \right)
    (a^\dagger a)_\xv   \nonumber\\
&& +\frac{1}{2}(1-\cos\thx)\nabla\phx \Jv_z(\xv)
   \left. +\frac{i}{2}\sum_{\pm}e^{\mp i\phx}
  (\mp\nabla\thx+i\sin\thx\nabla\phx) \Jv_\pm(\xv) \right],
\label{HI}
\eeqa
where $\Jv_\alpha (\xv)$ $(\alpha=\pm,z)$ are the spin currents of the
electron;
\beq
\Jv_\alpha (\xv,\tau)\equiv -\frac{i}{2m}[(a^\dagger\sigma_\alpha \nabla a)
- (\nabla a^\dagger\sigma_\alpha a)] ,
\label{curre}
\eeq
with $\sigma_\pm\equiv\sigma_x\pm i\sigma_y$.

The total Lagrangean written in the local frame is therefore given by
\beq
L=\sum_{\kv\sigma} a^\dagger_{\kv\sigma}(\partial_\tau+\epsilon_{\kv\sigma})
  a_{\kv\sigma}
   + \frac{U}{2}\int d^3x M^2 +H_{\rm int} ,
  \label{L2}
\eeq
where the electron energy is now spin dependent due to the background
magnetization (due to the term of eq.(\ref{int})):
\beq
\epsilon_{{\kv}\pm}\equiv \frac{{\kv}^2}{2m}\mp UM-\epsilon_{\rm F}.
\label{ener}
\eeq

\subsection{Effective action}
In order to derive the effective action for the description of the dynamics
of
the magnetization, one has to integrate out the electron degree of
freedom, for which we treat the interaction term $H_{\rm int}$
perturbatively.
This is reasonable since the variation of the magnetization vector is
generally very slow compared to that of electron.
For the case of a domain wall with thickness $\lambda$,
this perturbative treatment corresponds to the expansion in terms of
$(\kfu\lambda)^{-1}$, $\kfu$ being the fermi momentum of the majority spin.
The partition function is therefore written as
\beqa
\lefteqn{ Z= \int {\cal D}c^\dagger{\cal D}c{\cal D}M{\cal D}\nv
\delta(\nv^2-1)       }\nonumber\\
&&    \times \exp\left[-\int_0^\beta d\tau
\left( \sum_{\kv\sigma}a^\dagger_{\kv\sigma}
(\partial_\tau+\epsilon_{\kv\sigma})a_{\kv\sigma}
+\frac{U}{2}\int d^3 x M^2 +\Huf \right) \right]
\nonumber\\
&&   \times \left[1-\int_0^\beta d\tau H_{\rm int}+\half \int_0^\beta
 \int_0^\beta d\tau d\tau'
H_{\rm int}(\tau)H_{\rm int}(\tau') +O((H_{\rm int})^3)\right]
\nonumber\\
&&\simeq \int {\cal D}M{\cal D}\nv \delta(\nv^2-1)\exp\left[
{\rm tr}\ln(\partial_\tau+\epsilon_{\kv\sigma})-\beta\frac{U}{2}\int d^3xM^2
\right]
\nonumber\\
&& \times \exp\left[-\int_0^\beta d\tau <H_{\rm int}>
+\half \int_0^\beta\int_0^\beta d\tau d\tau' <H_{\rm int}(\tau)H_{\rm int}
(\tau') > \right]
\nonumber\\
&&\equiv \int {\cal D}M{\cal D}\nv \delta(\nv^2-1)
\exp\left[ -S_{\rm eff}(\theta,\phi,M) \right].
\eeqa
The brackets indicate the expectation values evaluated for the Lagrangean of
the electron $L_0\equiv \sum_{\kv\sigma}a^\dagger_{\kv\sigma}
(\partial_\tau+\epsilon_{\kv\sigma})a_{\kv\sigma}+\Huf$.
$\Huf$ is the Coulomb interaction, which contains processes with
finite momentum transfer only, and is given by
\beq
\Huf\equiv -\frac{U}{2V}\sum_{\qv\neq0}\sum_{\kv\kv'}
(a^\dagger_{\kv+\qv} \sigma_z a_\kv )(a^\dagger_{\kv'-\qv} \sigma_z a_{\kv'}
).
\label{huf}\eeq
This is the term which has not been taken into account in the determination
of the magnitude of
the magnetization, which has been assumed to be uniform.
The effective action for the magnetization is thus obtained as
\beq
S_{\rm eff}(\theta,\phi,M)=S_{\rm MF}(M)+\Delta S(\theta,\phi,M),
\label{Seff}
\eeq
where $S_{\rm MF}$ is the well-known mean field action of ferromagnet;
\beqa
\lefteqn{ S_{\rm MF}\equiv
-{\rm tr}\ln(\partial_\tau+\epsilon_{\kv\sigma})+\beta\int d^3x\frac{U}{2}
M^2
}\nonumber\\
&&=\beta \int d^3x\left[\frac{U}{2}M^2 - \frac{\epsilon_{\rm F}
(2m\epsilon_{\rm F})^{\frac{3}{2}}}{15\pi^2}
 \left\{\left(1+\frac{U}{\epsilon_{\rm F}}M\right)^{\frac{5}{2}}
 +\left(1-\frac{U}{\epsilon_{\rm F}}M\right)^{\frac{5}{2}}
 \right\}\right].  \label{Smf}
\eeqa
The dynamics of $(\theta,\phi)$ is determined by the second term in
eq.(\ref{Seff}), $\Delta S$,
\beqa
\Delta S&\equiv& \int_0^\beta d\tau <H_{\rm int}>
-\half \int_0^\beta d\tau\int d\tau' <H_{\rm int}(\tau)H_{\rm int}(\tau') >
\nonumber\\
&\equiv& \DSone+\DStwo.
\eeqa
The term $\DSone$ is the first order contribution, and is instantaneous
({\it i.e.}, depends only on a single time):
\beq
\DSone(\theta,\phi,M)=\int d\tau \int d^3x \left[ i\frac{M}{2}\dot{\phi}
(1-\cos\theta)
 +\frac{n}{4m}\left(\half (\nabla\theta)^2+(1-\cos\theta)(\nabla\phi)^2
 \right) \right],
\label{ds1}
\eeq
where $n\equiv <(a^\dagger a)_\xv>$ is the electron number per unit volume.
The terms in the second order are non-local in time.
Up to the first derivative in time, $\partial_\tau$, and second derivative
 in space, $\nabla^2$, it is given as
\beqa
\lefteqn{  \DStwo(\theta,\phi,M)=
  -\frac{1}{8}\int\! d\tau \! \int \! d\tau'\int d^3x \int d^3x' \sum_{ij}
 } \nonumber\\
&&
 \times \left\{
 \left[(1-\cos\theta)\nabla_i\phi \right]_{\tau\xv}
 \left[(1-\cos\theta)\nabla_j\phi \right]_{\tau'\xv'}
 <\Jv_z^i(\tau,\xv) \Jv_z^j(\tau',\xv')>
 \right. \nonumber\\
&&
  + \left.  2 e^{- i(\phi_{\tau\xv}-\phi_{\tau'\xv'})}
    (\nabla_i\theta-i\sin\theta\nabla_i\phi)_{\tau\xv}
    (\nabla_j\theta+i\sin\theta\nabla_j\phi)_{\tau'\xv'}
  <\Jv_+^i(\tau,\xv) \Jv_-^j(\tau',\xv')>
  \right\}.  \nonumber\\
&& \label{DS}
\eeqa

In the further calculations, the variable $M$ is approximated by its mean
field value $\Ms$ determined by the stationary condition of
$\delta S_{\rm MF}(\Ms)/\delta \Ms=0$, {\it i.e.,}
\beq
\Ms=\frac{(2m\epsilon_{\rm F})^{\frac{3}{2}}}{6\pi^2}
\left\{\left(1+\frac{U}{\epsilon_{\rm F}}\Ms\right)^{\frac{3}{2}}
 -\left(1-\frac{U}{\epsilon_{\rm F}}\Ms\right)^{\frac{3}{2}}
 \right\} . \label{MF}
\eeq
The fluctuation $\delta M$ of $M$ around $\Ms$ costs an energy of
\beq
\Delta E\simeq \int d^3x \frac{m_M^2}{2}(\delta M)^2,
\eeq
where the mass squared $m_{M}^2$ is given by the curvature of $S_{\rm MF}$
at $M=\Ms$ as
\beq
m_{M}^2=U\left[ 1-\frac{(2m\epsilon_{\rm F})^{\frac{3}{2}}}{4\pi^2}
\frac{U}{\epsilon_{\rm F}}
\left( \left(1+\frac{U}{\epsilon_{\rm F}}\Ms\right)^{\half}
+\left(1-\frac{U}{\epsilon_{\rm F}}\Ms\right)^{\half}
\right) \right] . \label{Mfl}
\eeq
For a strong ferromagnet with $\Ms a^3\gsim 0.1$ ($a$ being the lattice
constant), it is of order
of $m_M^2\gsim 0.01\times\ef\simeq100$K (see Fig. \ref{FIGMF}), and hence the
fluctuation $\delta M$ can be neglected for the study of low energy behavior.
In terms of $\Ms$ the fermi momenta of electrons with spin up and down are
given as follows by use of eq.(\ref{ener})
\beq
k_{{\rm F}\pm}\equiv (2m \epsilon_{\rm F})^{\half}\left
(1\pm \frac{U}{\epsilon_{\rm F}}\Ms\right)^{\half}. \label{kfdef}
\eeq
Equations (\ref{MF}) and (\ref{kfdef}) are equivalent to following equations,
\beqa
n&=&\frac{1}{6\pi^2}(\kfu^3 + \kfd^3)  \label{nn} \\
\Ms&=&\frac{1}{6\pi^2}(\kfu^3 - \kfd^3) .  \label{nm}
\eeqa
The behaviors of quantities $(\Ms a^3)$, $(\kfs a)$ and
$(m_{M}^2/\ef a^3)$ as a function of
$(U/\epsilon_{\rm F}a^3)$ (with the electron number per site ($na^3$) being
fixed to be unity) are shown in Fig. \ref{FIGMF}.
As is well known, the ferromagnetism appears for $(U/\epsilon_{\rm F}a^3)
 >(2/3)\nzero^{-1}$ where
$\nzero\equiv (2m\epsilon_{\rm F})^{3/2}a^3/(3\pi^2)$.
For $(U/\epsilon_{\rm F}a^3)\geq1$, the magnetization saturates and $\kfd$
vanishes. These are the results of the Hartree-Fock theory, which describes
the essential features of itinerant ferromagnetism but should not be taken
literally in comparison with the actual experiments.

The expectation values of  electron  spin currents $<\Jv\Jv>$'s  in $\DStwo$,
eq.(\ref{DS}), are written as
\beq
<\Jv_\alpha^i(\xv\tau) \Jv_\beta^j(\xv'\tau')>\equiv \frac{1}{\beta}\sum_\ell
e^{i\omega_\ell(\tau-\tau')} \frac{1}{V}\sum_\qv e^{-i\qv\cdot(\xv-\xv')}
<\Jv_\alpha^i(\qv,\ell) \Jv_\beta^j(-\qv,-\ell)>, \label{JJfou}
\eeq
where $\alpha,\beta=\pm,z$, and
 $\oml$ being the thermal frequency, $\oml\equiv 2\pi\ell/\beta$.
The Fourier transform of the spin current is defined by
\beq
J_\alpha^i(\qv,\ell)\equiv \frac{1}{\sqrt{\beta}}\frac{1}{\sqrt{V}}
\sum_{\kv}\sum_{\omega_n}\frac{1}{m}\left(k+\frac{q}{2}\right)^i
(a_{\kv+\qv, n+\ell}^\dagger \sigma_\alpha a_{\kv,n}),
\eeq
with $\omega_n\equiv (2n+1)\pi/\beta$.
The correlation functions are calculated by the random phase approximation
(RPA) to the Coulomb interaction, $\Huf$ (eq.(\ref{huf})).
Hence the correlation $<\Jv_+\Jv_->$ is evaluated, as
depicted in Fig. \ref{FIGRPA}(a) diagramatically, to be
\beq
   <\Jv_+^i(\qv,\ell) \Jv_-^j(-\qv,-\ell)>  = \Cpm^{ij}(\qv,\ell)
+\Cpm^{i}(\qv,\ell)\Cpm^{j}(\qv,\ell)\frac{2U}{1-2U\chi^0_{+-}(\qv,\ell)} ,
\label{JJpm}
\eeq
where
\beqa
\Cpm^{ij}(\qv,\ell)&\equiv& \frac{1}{V}\frac{1}{4m^2}
 \sum_{\kv}(2k+q)^i(2k+q)^j\frac{\fkd-\fkqu}{\ekqu-\ekd-i\oml} \nonumber\\
\Cpm^{i}(\qv,\ell)&\equiv& \frac{1}{V}\frac{1}{2m}
 \sum_{\kv}(2k+q)^i\frac{\fkd-\fkqu}{\ekqu-\ekd-i\oml} ,\label{c+-}
\eeqa
and the irreducible spin susceptibility $\chi^0_{+-}$ is given by
\beq
\chi^0_{+-}(\qv,\ell)\equiv\frac{1}{V}\sum_{\kv}\frac{\fkd-\fkqu}
{\ekqu-\ekd-i\oml}  .\label{chi+-}
\eeq
Here $f_{\kv\sigma}$ is the fermi distribution function
\beq
f_{\kv\sigma}\equiv\frac{1}{e^{\beta({\kv^2}/{2m}-\efs)}+1},
\eeq
with the fermi energy $\efs\equiv \epsilon_{\rm F}\pm U\Ms$.
Similarly, as seen from from Fig. \ref{FIGRPA}(b),
\beqa
<\Jv_z^i(q,\ell) \Jv_z^j(-q,-\ell)> &=& \sum_\sigma\left[C_{\sigma}^{ij}
(\qv,\ell)
+C_{\sigma}^{i}(\qv,\ell)C_{\sigma}^{j}(\qv,\ell)\frac{U^2 \chi^0_{-\sigma}
(\qv,\ell)}{1-U^2\chi^0_{\uparrow}(\qv,\ell)\chi^0_{\downarrow}(\qv,\ell)}
\right]   \nonumber\\
&& +\frac{2U C_{\uparrow}^{i}(\qv,\ell)C_{\downarrow}^{j}(\qv,\ell)}
{1-U^2\chi^0_{\uparrow}(\qv,\ell)\chi^0_{\downarrow}(\qv,\ell)},
\eeqa
where
\beqa
C_{\sigma}^{ij}(\qv,\ell)&\equiv& \frac{1}{V}\frac{1}{4m^2}
 \sum_{\kv}(2k+q)^i(2k+q)^j\frac{\fks-\fkqs}{\ekqs-\eks-i\oml}  \nonumber\\
C_{\sigma}^{i}(\qv,\ell)&\equiv& \frac{1}{V}\frac{1}{2m}
 \sum_{\kv}(2k+q)^i\frac{\fks-\fkqs}{\ekqs-\eks-i\oml} ,
\eeqa
and
\beq
\chi^0_{\sigma}(\qv,\ell)\equiv\frac{1}{V}\sum_{\kv}\frac{\fks-\fkqs}
{\ekqs-\eks-i\oml}  .
\eeq

\subsection{Effective Heisenberg model (local part)}
The second order action $\DStwo$ is made up with a local part, that
contributes to the renormalization of the action
$\DSone$, and the non-local part describing the dissipative effect.
To see this, we rewrite $\DStwo$ as
\beqa
\lefteqn{  \DStwo(\theta,\phi)=
 - \frac{1}{8}\int\! d\tau \! \int \! d\tau' \frac{1}{\beta}\sum_\ell
e^{i\omega_\ell(\tau-\tau')} \sum_{ij} \frac{1}{V}\sum_\qv
 } \nonumber\\
&&
 \times  \left(
  \Phi_\qv^i(\tau)\Phi_\qv^j(\tau')^\dagger <\Jv_z^i(\qv,\ell)
 \Jv_z^j(-\qv,-\ell)>
 +2\Theta_\qv^i(\tau)\Theta_\qv^j(\tau')^\dagger
 <\Jv_+^i(\qv,\ell) \Jv_-^j(-\qv,-\ell)>\right),  \nonumber\\
&&
\eeqa
where
\beqa
 \Phi_\qv^i(\tau)&\equiv& \int d^3x
  e^{-i\qv\xv}(1-\cos\theta)\nabla_i\phi \label{Phidef}\\
 \Theta_\qv^i(\tau)&\equiv& \int d^3x
 e^{-i\qv\xv}(\nabla_i\theta-i\sin\theta\nabla_i\phi) .
 \label{Thetadef}
\eeqa
The correlation functions $<\Jv_\alpha^i(\qv,\ell) \Jv_\beta^j(-\qv,-\ell)>$
are diagonal in indexes $i$ and $j$
({\it i.e.,} $\propto \delta_{ij}$), and so $\DStwo$ is written as
\beqa
\lefteqn{  \DStwo(\theta,\phi)=
  \frac{1}{8}\int\! d\tau \! \int \! d\tau' \frac{1}{\beta}\sum_\ell
e^{i\omega_\ell(\tau-\tau')} \sum_{i}\frac{1}{V}\sum_\qv
 } \nonumber\\
& &
 \times   \left[
 \left( - \half \left( |\Phi_\qv^i(\tau)|^2+|\Phi_\qv^i(\tau')|^2\right)
 +|\Phi_\qv^i(\tau)-\Phi_\qv^i(\tau')|^2 \right)
 <\Jv_z^i(\qv,\ell) \Jv_z^i(-\qv,-\ell)>
\right. \nonumber\\
& & \left.
 +2\left( -\half \left( |\Theta_\qv^i(\tau)|^2+|\Theta_\qv^i(\tau')|^2
\right)
 +|\Theta_\qv^i(\tau)-\Theta_\qv^i(\tau')|^2 \right)
 <\Jv_+^i(\qv,\ell) \Jv_-^i(-\qv,-\ell)>\right]. \nonumber\\
& &
\eeqa
In terms that contain the field at a single time, such as
$|\Phi_\qv^i(\tau)|^2$, the integral over one time variable, $\tau'$, leads
to
$\delta_{\ell,0}$, and hence the action is devided into two parts;
\beqa
\lefteqn{  \DStwo(\theta,\phi)= -\frac{1}{8} \int\! d\tau \frac{1}{V}
 \sum_\qv\sum_i   } \nonumber\\
&&\times \left[ |\Phi_\qv^i(\tau)|^2  <\Jv_z^i(\qv,0) \Jv_z^i(-\qv,0)>
\right.
 \left.  +2|\Theta_\qv^i(\tau)|^2
<\Jv_+^i(\qv,0) \Jv_-^i(-\qv,0)> \right]
  \nonumber\\
&&
+  \frac{1}{8}\int\! d\tau \! \int \! d\tau' \frac{1}{\beta}\sum_\ell
e^{i\omega_\ell(\tau-\tau')} \sum_{i}\frac{1}{V}\sum_\qv
 \left[|\Phi_\qv^i(\tau)-\Phi_\qv^i(\tau')|^2
 <\Jv_z^i(\qv,\ell) \Jv_z^i(-\qv,-\ell)> \right.
\nonumber\\
&&
   \left.
  +2|\Theta_\qv^i(\tau)-\Theta_\qv^i(\tau')|^2 <\Jv_+^i(\qv,\ell)
\Jv_-^i(-\qv,-\ell)>\right]  \nonumber\\
&&\equiv \DStwo_{\rm loc}+\DStwo_{\rm dis} \label{DSLD}
{}.
\eeqa
We have defined the local part $\DStwo_{\rm loc}$ by the first two terms in
eq.(\ref{DSLD}) and $\DStwo_{\rm dis}$ by the last two terms.
In the local part, $\DStwo_{\rm loc}$,  the component of $\qv=0$ gives rise
to terms with second order derivative $\nabla ^2$ of $\theta$ and $\phi$;
\beqa
 \DStwo_{\rm loc}(\theta,\phi)&\simeq& -\frac{1}{8}\int\! d\tau
\int d^3x\sum_i
\left[ (1-\cos\theta)^2(\nabla_i\phi)^2  <\Jv_z^i(0,0) \Jv_z^i(0,0)>
\right.\nonumber\\
&& + \left.
2((\nabla_i\theta)^2+\sin^2\theta(\nabla_i\phi)^2)
<\Jv_+^i(0,0) \Jv_-^i(0,0)> \right].
\label{Slocren}
\eeqa
The terms arising from the finite $\qv$ component contains higher derivatives
and can be neglected.
Concerning the $q=0$ part of the current correlation functions,
 only the irreducible (one-loop) diagram
contributes  due to the current, $2\kv+\qv$, at the vertex,
and the correlation functions are calculated as
\beqa
\lim_{q\rightarrow0}<\Jv_+^i(\qv,0) \Jv_-^j(-\qv,0)>& =&
       \frac{(\kfu^{5}-\kfd^5)}{60\pi^2m^2U\Ms} \delta_{ij}   \nonumber\\
\lim_{q\rightarrow0}<\Jv_z^i(\qv,0) \Jv_z^j(-\qv,0)>& =&
    \frac{n}{m}\delta_{ij}.
\eeqa
Consequently the local part ($\DSone$, eq.(\ref{ds1}), and
 $\DStwo_{\rm loc}$)
of the effective action $\Delta S$ is given by
\beq
\Delta S_{\rm loc}=\int d\tau \int {d^3\xv}
  \left[ i\frac{S}{a^3}\dot{\phi}(1-\cos\theta)+
   \frac{JS^2}{2}\left( (\nabla\theta)^2+\sin^2\theta(\nabla\phi)^2 \right)
  \right], \label{Sloc}
\eeq
where the spin is $S\equiv \Ms a^3/2$ and the exchange coupling or the spin
stiffness $J$ ([J/m] in MKSA unit) is expressed by the parameters of
itinerant
electrons as
\beq
J\equiv\frac{n}{ma^6\Ms^2}\left[1-\frac{(\kfu^{5}-\kfd^5)}{30\pi^2 mnU\Ms}
\right] . \label{Jdef}
\eeq
The action eq.(\ref{Sloc}) is of the same form as that of the ferromagnetic
Heisenberg model with $S=\Ms a^3/2$, as has already been
indicated\cite{Pra,RS}.
Hence, there is no formal difference from the case of an
insulator in the local part of the effective action, but
$S$ in the present itinerant system is a continuous variable in contrast
to the Heisenberg model.
The behavior of the exchange energy $(J\Ms^2 a^7/\epsilon_{\rm F})$ is
plotted as a function of
$(U/\epsilon_{\rm F}a^3)$ in Fig. \ref{FIGMF}.
\subsection{Dissipative part (non-local part)}
We now go on to the non-local part $\DStwo_{\rm dis}$ defined by
eq.(\ref{DSLD}),
where a characteristic feature of the itinerant model is contained.
In fact, this part represents the dissipative effect from the itinerant
electrons.
To look into this effect that dominates at low energy, the
analytical continuation
 to the real frequency $i\oml\rightarrow\omega\pm i0$ is convenient\cite{ATF}.
Let us take, as an example, the term in $\DStwo_{\rm dis}$ proportional
to $<\Jv_+\Jv_->$.
It is
\beq
  \frac{1}{4}\int\! d\tau \! \int \! d\tau' \frac{1}{\beta}\sum_\ell
 e^{i\oml(\tau-\tau')}
 \frac{1}{V}\sum_{\qv} \sum_{i}|\Theta_\qv^i(\tau)-\Theta_\qv^i(\tau')|^2
     <\Jv_+^i(\qv,\ell) \Jv_-^i(-\qv,-\ell)> .\label{DSexamp}
\eeq
In this expression, the summation over the thermal frequency $\oml$ is
rewritten using the contour integration  as
\beqa
\lefteqn{ \frac{1}{\beta}\sum_{\ell}e^{i\oml(\tau-\tau')}
<\Jv_+(\qv,\ell) \Jv_-(-\qv,-\ell)> }\nonumber\\
&& =\frac{1}{2\pi i}\int_C dz
 \left( \frac{e^{z(\tau-\tau')}}{e^{\beta z}-1}\theta(\tau-\tau')
  +\frac{e^{z(\tau-\tau')}}{1-e^{-\beta z}}\theta(\tau'-\tau)  \right)
  <\Jv_+(\qv) \Jv_-(-\qv)>|_{i\oml\equiv z}, \nonumber\\
&& \label{cont}
\eeqa
where the coutour $C$ is along the imaginary axis (see Fig. \ref{FIGcont}).
Deforming the path $C$ into $C'$ shown in Fig. \ref{FIGcont}, the above
expression reduces to
\beqa
&&\rightarrow {\rm P}\int_{-\infty}^\infty \frac{d\omega}{\pi}
\left( \frac{e^{\omega(\tau-\tau')}}{e^{\beta\omega}-1}\theta(\tau-\tau')
  +\frac{e^{\omega(\tau-\tau')}}{1-e^{-\beta\omega}}\theta(\tau'-\tau)
 \right)
{\rm Im}<\Jv_+(\qv) \Jv_-(-\qv)>|_{\omega+i0}  , \nonumber\\
&&  \label{AC}
\eeqa
where P denotes the principal value, and
the correlation function $<\Jv(q)\Jv(-q)>|_{\omega+i0}$ in the right hand
side is defined by
$<\Jv(q,\ell)\Jv(-q,-\ell)>$ with $i\oml$ replaced by $\omega+i0$.
We thus need to calculate only the imaginary part of the correlation
functions
as a function of a real frequency in order to investigate the effect of
dissipation.

The important contribution is that of the ohmic dissipation,  which is due to
the $\omega$-linear term in Im$<\Jv(\qv) \Jv(-\qv)>|_{\omega+i0}$ as
$\omega\rightarrow0$.
With this ohmic dissipation, the above expression eq.(\ref{AC}) reduces to
\beq
\rightarrow \left(\frac{\pi}{\beta}\right)^2\frac{1}{\sin^2\frac{\pi}{\beta}
\tau}\lim_{\omega\rightarrow0}
\left(\frac{{\rm Im}<\Jv_+(\qv) \Jv_-(-\qv)>|_{\omega+i0}}{\pi\omega}\right),
\label{ohm}
\eeq
and its $\tau$ dependence becomes $\tau^{-2}$ at zero temperature.

After straightfoward algebra, we see that
the functions $C$ and $\chi^0$  at small $\omega$ needed for the calculation
of $<\Jv_+\Jv_->$ are obtained as follows.
\beqa
\chi^0_{+-}(\qv,\omega+i0)&=& A_q+i\omega B_q \theta_{\rm st}(q)+O(\omega^2)
\nonumber\\
\Cpm^{i}(\qv,\omega+i0)&=& \left[ -\frac{1}{q}\Ms(1-2UA_q)
   +i\omega \frac{(\kfu^2-\kfd^2)}{2mq} B_q \theta_{\rm st}(q) \right]
  \delta_{i,1}
   +O(\omega^2) \nonumber\\
{\rm Im}\Cpm^{ij}(\qv,\omega+i0)&=&\omega \frac{(\kfu^2-\kfd^2)^2}{4m^2q^2}
 B_q \theta_{\rm st}(q)
\delta_{i,1}\delta_{j,1}+O(\omega^2),  \label{Imc}
\eeqa
where we have chosen the direction of $\qv$ along the $x$ axis.
The functions $A_q$ and $B_q$ are given by
\beqa
 A_q&=&\frac{m}{8\pi^2}\left[(\kfu+\kfd)
 \left(1+\frac{(\kfu-\kfd)^2}{q^2}\right)
 \right. \nonumber\\
&& +  \left. \frac{1}{2q^3}((\kfu+\kfd)^2-q^2)(q^2-(\kfu-\kfd)^2) \ln
  \left| \frac{q+(\kfu+\kfd)}{q-(\kfu+\kfd)} \right|  \right]  \nonumber\\
B_q &= & \frac{m^2}{4\pi |q|}.  \label{B}
\eeqa
The ohmic dissipation appears only for the restricted region of
momentum variable specified by the function
$\theta_{\rm st}(q)$:
\beqa
\theta_{\rm st}(q)\equiv \left\{ \begin{array}{ccc} 1  & \mbox{  }  &
  (\kfu-\kfd)<|q|<(\kfu+\kfd)   \\
                                            0  &   &
  {\rm otherwise} \end{array} \right.\label{st}
\eeqa
Thus ohmic dissipation is seen to be due to the Stoner excitation,
which is a gapless excitation that flips a spin of an electron at the
fermi surface.
It is seen from eq.(\ref{B}) that these procedures of taking out the
$\omega$-linear term is the expansion in terms of
$(\omega/\epsilon_{F\uparrow})(\kfu/(\kfu\pm\kfd))$.
Collecting these expressions, we obtain
\beq
{\rm Im}<\Jv^1_+(\qv) \Jv^1_-(-\qv)>|_{\omega+i0}=
  \pi\omega\frac{(\kfu^2-\kfd^2)^2}{4\pi^2|q|^3}\theta_{\rm st}(q) +
  O(\omega^3).  \label{Im}
\eeq
In the following analysis, we consider a planar wall with a spin
configuration
changing only in $x$-direction in space, and the spins lying in the
$yz$-plane; {\em i.e.,} $\phi=\pi/2$ (see Fig. \ref{FIGDW}(a)).
In this  case, $\Phi_\qv^i=0$ and thus only the $<\Jv_+ \Jv_->$ term  in
eq.(\ref{DSLD}) is relevant.
Therefore the effect of dissipation at zero temperature is finally given as
\beqa
\lefteqn{ \Delta S_{\rm dis}=
\frac{(\kfu^2-\kfd^2)^2 }{32\pi^2 }\int d\tau \int d\tau'
  \frac{1}{(\tau-\tau')^2}  }\nonumber\\
&&\times
\int\frac{d^3q}{(2\pi)^3} |\Theta_\qv^1(\tau)- \Theta_\qv^1(\tau')|^2
 \frac{1}{|q|^3} \theta_{\rm st}(q),  \label{Sdis}
\eeqa
where  the momentum integration is over the region of Stoner excitation
and the form factor of the wall $\Theta_\qv^1(\tau)$ is
\beq
\Theta_\qv^1(\tau)=\int d^3\xv e^{-i\qv\xv}\nabla_x \theta(\xv,\tau).
\label{Thetdef}
\eeq
The total action for the direction of the magnetisation vector is
$\Delta S_{\rm loc}+\Delta S_{\rm dis}$.
Since $\Delta S_{\rm dis}$ is positive definite, the tunneling rate is
seen to be always reduced by this dissipation effect within the present
semi-classical approximation.
In the case of a complete ferromagnetic case, $(U/\epsilon_{\rm F}a^3)\geq1$,
$\kfd=0$ and then $\Delta S_{\rm dis}$ is vanishing.

\section{Effects of Dissipation on MQT of a Domain Wall}
\label{SECdw}
In the previous section, we have derived based on the Hubbard model an
effective action $\Delta S_{\rm loc}+\Delta S_{\rm dis}$
describing the slowly varying direction of the magnetization vector.
By use of  this effective action, we will investigate the quantum tunneling
of a domain wall in the case as shown in Fig.\ref{FIGDW}(a).
To do this, we need first to describe the domain wall motion in terms of
the tunneling variable, the coordinate of the the wall $Q$.
The dissipative effect $\Delta S_{\rm dis}$ is to be estimated by
use of the variable $Q$.
\subsection{MQT of a domain wall}
\label{SECdw1}
Let us first study the local part of the action $\Delta S_{\rm loc}$.
The analysis in this subsection is based on the discussion by
Stamp\cite{Sta,CIS,SCB}.
For the domain wall to exist, anisotropy energy is needed, which we
 shall add phenomenologically to the action $\Delta S_{\rm loc}$ as the
 energy $-K$[J/m$^3$] in the $z$-direction as an easy axis and
$+K_{\perp}$ in the $x$-direction as a hard axis.
We consider the case of strong anisotropy; $K,K_{\perp}\gg K_d$,
where $K_d\equiv \mu_0(\hbar\gamma)^2/a^6$ is the magnetostatic
energy due to demagnetization
($\gamma=e/(2m)$ is the gyromagnetic ratio
and $\mu_0$ is the magnetic permeability of a free space).
The calculation below also applies to a ferromagnet with uniaxial
anisotropy $-K$ in $z$-direction, if one replaces $K_{\perp}\rightarrow K_d$,
as will be discussed in Appendix \ref{demag}.
The effect of the anisotropy energy on the dissipation due to the itinerant
electron is neglected, since the correction would be small by the order of
$(Ka^3/U\Ms)\simeq 10^{-3}$ (for $Ka^3\simeq10$K, $\Ms a^3\gsim0.1$).

The action we deal with is therefore given by the following
 (including $\Delta S_{\rm dis}$)
\beqa
 S(\theta, \phi)&=&\int d\tau \int {d^3 \xv} \left[
  i\frac{\Spin}{a^3}\dot{\phi}(1-\cos\theta)+
   \frac{J\Spin^2}{2}\left( (\nabla\theta)^2+\sin^2\theta(\nabla\phi)^2
\right)
\right.
\nonumber\\
&&-\left.\frac{K}{2}\Spin^2\cos^2\theta +\frac{K_{\perp}}{2}
 \Spin^2\sin^2 \theta\cos^2\phi \right] +\Delta S_{\rm dis}
. \label{Smag}
\eeqa
The spin, $S=\Ms a^3/2$, and the exchange coupling $J$[J/m] are determined
microscopically by eqs.(\ref{MF}) and (\ref{Jdef}).
In the absence of the transverse anisotropy, $K_{\perp}$,
the domain wall cannot tunnel.
Actually, without this term, the $z$ component of spin $S_z(\xv)$ is
conserved ({\it i.e.,} commutes with Hamiltonian) at each site $\xv$.
This fact is expressed in eq.(\ref{Mwall}) below as the divergence of
the domain wall mass as $K_\perp\rightarrow 0$.

Without dissipation, $\Delta S_{\rm dis}=0$, the action eq.(\ref{Smag}) has
a classical solution of a planar domain wall, where the spin configuration
changes only in one direction, which we choose as $x$-axis, as depicted in
Fig.\ref{FIGDW}(a). The position of the wall is  at $x=Q$.
Because of the anisotropy, the spin points to the $\pm$z direction at
infinity and it rotates
in the $yz$-plane (due to transverse anitsotropy energy that unfavors the
$x$-direction) around $x=Q$.
For a wall moving with a small velocity $\dot{Q} \ll c$, where
$c\equiv K_\perp \lambda Sa^3/\hbar$, the classical solution is given by
\beqa
\cos\theta(\xv, \tau) &=& \tanh \frac{x-Q(\tau)}{\lambda}, \nonumber\\
\cos\phi(\xv,\tau)&\simeq&i\frac{\dot{Q}}{c} \ll 1 ,\label{DWT}
\eeqa
or in terms of the vector $\nv$,
\beqa
\nv(\xv,\tau)=\left(\begin{array}{c}
\left[\cosh \frac{x-Q(\tau)}{\lambda}\right]^{-1} i\frac{\dot{Q}}{c} \\
      \left[{\cosh \frac{x-Q(\tau)}{\lambda}} \right]^{-1}
                  \left[1+\frac{1}{2}\left(\frac{\dot Q}{c}\right)^2\right]
   \\
                        \tanh \frac{x-Q(\tau)}{\lambda}   \end{array}\right)
+O\left(\frac{\dot Q}{c}\right)^3 .
\label{DW}
\eeqa
The width of the wall is given by
\beq
\lambda\equiv \sqrt{\frac{J}{K}} .
\eeq

The dynamics of the wall coordinate $Q(\tau)$ is determined by the following
macroscopic considerations.
For the configuration of eq.(\ref{DWT}), the Lagrangean in eq.(\ref{Smag})
(terms in square bracket) is written in terms of $Q$ as (adding irrelavant
constant)
\beqa
L_w&=&S^2\left[K-\frac{K_\perp}{2}\left(\frac{\dot Q}{c}\right)^2\right]
\int d^3 x \frac{1}{\cosh^2\frac{x-Q}{\lambda}}\nonumber\\
&=&2\Awall\lambda KS^2 -\frac{1}{2}M_{\rm w}\dot Q^2, \label{LQ}
\eeqa
where $\Awall$ is the area of the wall and
\beq
M_{\rm w}\equiv\frac{2\Awall}{K_\perp\lambda}.     \label{Mwall}
\eeq
The first term in eq.(\ref{LQ}) is the static energy of the domain wall while
the second term is the kinetic energy of the wall
(note $-\dot Q^2=(dQ/dt)^2$ in real time, $t\equiv i\tau$), and hence
$M_{\rm w}$ represents the mass of a domain wall.
The pinning potential energy for $Q$ produced by the defect is assumed to
be short ranged\cite{Sta}
whose strength is to be related to the coercive field $H_c$ of the system.
The magnetic field $H$ produces a linear potential $\propto HQ$ and thus
makes the pinning center $Q=0$ metastable (see Fig. \ref{FIGpot}).
In the experimental situations, the field is set very close to the
coercive field $H_c$,
where the energy barrier vanishes,  for the tunneling rate to be large enough.
In this small energy barrier case, the total potential for $Q$ produced
by the defect and magnetic field is well approximated
by a sum of  harmonic and cubic potential\cite{SCB}.
\beq
V(Q)=\half M_{\rm w} \omega_0^2 Q^2\left(1-\frac{Q}{Q_0}\right), \label{pot}
\eeq
as shown in Fig. \ref{FIGpot}.
Here the attempt frequency $\omega_0$ and the tunneling barrier width $Q_0$
is
expressed in terms of macroscopic parameters in the leading order of
$\epsilon\equiv(H_c-H)/H_c$ as\cite{SCB}
\beqa
\omega_0 &\simeq& \mu_0 \frac{(\hbar\gamma)^2 S}{a^3}\sqrt{h_c}
\epsilon^{\frac{1}{4}} \simeq
 10^{11}\times \sqrt{h_c}\epsilon^{\frac{1}{4}} \,\,\, ({\rm Hz})
 \simeq 5\times \sqrt{h_c}\epsilon^{\frac{1}{4}} \,\,\, ({\rm K})
\label{omegazero}\\
Q_0 &=& \sqrt{\frac{3}{2}}\sqrt{\epsilon}\lambda ,\label{Qzero}
\eeqa
where $h_c\equiv H_c/(2\hbar\gamma S /a^3)$ is the ratio of the coercive
field to the magnetic moment per unit volume.
The typical values of these parameters are $h_c\simeq 10^{-4}\sim10^{-2}$ and
$\epsilon \lsim 1$.
A factor of $10^{11}$ (Hz) in the expression of $\omega_0$
arises from the magnetostatic energy due to the magnetization
of $(2\hbar\gamma S/a^3)\simeq 10^6$[A/m] ({\it i.e.,} $S\simeq O(1)$).
Note that the barrier width $Q_0$ is much smaller than the domain wall
width $\lambda$ for the case
of shallow potential ({\it i.e.,} small $\epsilon$).
The barrier height is proportional to the number of spins in the wall
$N\equiv\Awall\lambda/a^3$, and is expressed as
\beq
U_{\rm H}\simeq \mu_0 \frac{(\hbar\gamma S)^2 }{a^3}N h_c
\epsilon^{\frac{3}{2}} \simeq
 5\times N{h_c}\epsilon^{\frac{3}{2}} \,\,\, ({\rm K})\label{barrier}.
\eeq

{}From these considerations, the dynamics of $Q$ is determined by the
Lagrangean
(subtracting a constant)
\beqa
L(Q)&\equiv& L_w+V(Q) \nonumber\\
 &=&  \half M_{\rm w}\dot{Q}^2
  +  \half M_{\rm w} \omega_0^2 Q^2\left(1-\frac{Q}{Q_0}\right) .
\eeqa
The tunneling rate of the variable $Q$ from the metastable state $Q=0$ of
the potential $V(Q)$ is estimated
using the bounce solution of $L(Q)$, which describe $Q$ travelling from $0$
at $\tau=-\infty$
to $Q_0$ at $\tau=0$ and goes back to $Q=0$ at $\tau=\infty$\cite{CC}
(see Fig. \ref{FIGpot}).
It is explicitly given as
\beq
Q(\tau)=Q_0\frac{1}{\cosh^2\frac{\omega_0\tau}{2}}. \label{bounce}
\eeq
The exponent of the tunneling rate without dissipation is given by the
value of action
$B\equiv \int d\tau L(Q)$ estimated along the bounce solution.
Taking into account the prefactor, the tunneling rate without dissipation is
estimated as
\beq
\Gamma_0=Ae^{-B} , \label{rate0}
\eeq
where
\beqa
A&\simeq& 10^{11} h_c^{\frac{3}{4}}N^{\half}\epsilon^{\frac{7}{8}}
\,\,\,({\rm Hz})
\nonumber\\
B&\simeq& h_c^\half N \epsilon^{\frac{5}{4}}.\label{AB}
\eeqa
The factor of $\epsilon^{5/4}$ arises
from the barrier height and width in the small $\epsilon$ limit
as seen by the WKB approximation;
$\Gamma_0 \propto$(barrier height)$^{1/2}
\times Q_0\propto \epsilon^{3/4}\epsilon^{1/2}$.
Because $h_c$ and $\epsilon$ are small, MQT of a domain wall is observable
even for very large number of spins if the dissipation is negligible.
For example, for $h_c=10^{-4}$ and $N=10^{6}$, MQT is observable
({\it e.g.,} $\Gamma_0 \gsim 10^{-4}$Hz) by
setting the magnetic field within 1\% of the coercive field ({\it i.e.},
$\epsilon\lsim10^{-2}$)(see dashed line in Fig.\ref{FIGgamma}).
Instead, $\epsilon\lsim2\times10^{-4}$ is needed for $N=10^8$.

The crossover temperature $T_{\rm co}$ from the thermal activation to the
quantum tunneling is estimated by the relationship $B=U_{\rm H}/T_{\rm co}$
where $U_{\rm H}$ is the barrier height of the potential.
For the present problem, the crossover temperature is roughly the same as the
attempt frequency $\omega_0$;
\beq
T_{\rm co}\simeq 5\times \sqrt{h_c} \epsilon^{\frac{1}{4}}\,\,\,({\rm K}).
\label{Tc0}
\eeq
%
\subsection{Effect of dissipation}
Now we discuss the effect of dissipation due to itinerant electrons,
$\Delta S_{\rm dis}$.
Within the calculation based on the bounce solution, the tunneling rate
eq.(\ref{rate0}) is reduced by dissipation to be\cite{CL}
\beq
\Gamma=Ae^{-(B+\Delta {S}_{\rm dis})} =\Gamma_0 e^{-\Delta {S}_{\rm dis}} .
\label{rate}
\eeq
Here, to the lowest order in dissipation, the exponent
$\Delta {S}_{\rm dis}$ is evaluated as the value of the action
eq.(\ref{Sdis})
 estimated along the domain wall solution eq.(\ref{DWT}) with
$Q(\tau)$ given by the bounce eq.(\ref{bounce}).
For the present planar domain wall solution eq.(\ref{DWT}), we see $\nabla
\phi=0$ and
\beq
\nabla_j\theta(\xv)=
 -\frac{1}{\lambda}\frac{1}{\cosh\frac{x-Q(\tau)}{\lambda}}\delta_{j,1} .
\eeq
The Fourier transformation of this quantity, $\Theta_\qv^j$ in
eq.(\ref{Thetdef}), is
\beq
\Theta_\qv^j(\tau)= -\pi A_w e^{-iq_1 Q(\tau)}\frac{1}{\cosh
\frac{\pi}{2}\lambda q_1}
\delta_{q_2,0}\delta_{q_3,0}\delta_{j1}, \label{thetbounce}
\eeq
where $q_i$'s are the $i$-th component of the momentum $\qv$.
The $\delta$-functions as regards to $q_2$ and $q_3$ arises  because the
configuration of the magnetization is uniform in the $yz$-plane.
$\Delta {S}_{\rm dis}$ thus reads as follows
\beqa
\lefteqn{ \Delta {S}_{\rm dis}= \frac{(\kfu^2-\kfd^2)^2}{16} \Awall \int
d\tau\int d\tau' \frac{1}{(\tau-\tau')^2}
 }\nonumber\\
&&\times \int\frac{dq_1}{2\pi}\left[1-\cos{q_1(Q(\tau)-Q(\tau'))}\right]
 \frac{1}{\cosh^2 \frac{\pi}{2}\lambda q_1} \frac{1}{|q_1|^3}
\theta_{\rm st}(q_1) \nonumber\\
&=&N\frac{(\kfu^2-\kfd^2)^2a^4}{4}\frac{1}{\lambda a}
\int d\tau\int d\tau' \frac{1}{(\tau-\tau')^2}  \nonumber\\
&&\times\int^{\kfu+\kfd}_{\kfu-\kfd}\frac{dq_1}{2\pi} \sin^2\frac{q_1}{2}
(Q(\tau)-Q(\tau'))
\frac{1}{\cosh^2 \frac{\pi}{2}\lambda q_1} \frac{ 1}{q_1^3} ,\label{DSdis}
\eeqa
where $N$ is the number of spins in the wall, $N\equiv \Awall \lambda/a^3$.

Because of the form  of the wall, $1/\cosh ^2 (\pi\lambda q/2)$,
$\Delta S_{\rm dis}$ can be
large if $(\kfu-\kfd)\lambda \lsim 1$.
As long as this inequality holds, the integral is dominated by
$q\lsim\lambda^{-1}$ and thus,
noting eq.(\ref{Qzero}) with $\epsilon\lsim1$, the sine function may be
replaced by its argument.
Therefore
\beqa
\Delta S_{\rm dis}&\simeq& N\frac{(\kfu^2-\kfd^2)^2a^4}{32\pi}
\frac{Q_0^2}{\lambda a}
\int d\tau\int d\tau' \frac{(Z(\tau)-Z(\tau'))^2}{(\tau-\tau')^2}
\nonumber\\
&&
\times \int^{(\kfu+\kfd)\frac{\pi}{2}\lambda}_{(\kfu-\kfd)\frac{\pi}{2}
\lambda}
\frac{dx}{x} \frac{1}{\cosh^2 x}  , \label{DS3}
\eeqa
where $Z(\tau)\equiv Q(\tau)/Q_0=1/\cosh^2 (\omega_0 \tau/2)$.
In evaluating the time integral, we note that the bounce solution
$Z(\tau)$ is localized within $|\tau|\lsim \omega_0^{-1}$.
The integration must be cut off at short time $\sim \omega_0^{-1}$ as shown
originally by Kagan and Prokof'ev\cite{KP,Iso}.
The physical meaning of this cutoff may be considered as follows\cite{KP}.
In the calculation of $\Delta S_{\rm dis}$, we have made use of the bounce
solution, eq.(\ref{bounce}), with zero energy.
In other words, we have neglected the excited states of variable $Q$.
This approximation would be valid only
for small energy transfer $\omega\lsim \omega_0$ in the
current correlation function $<\Jv\Jv>$.
The high energy excitation of electron $\omega\gsim \omega_0$ produces
only  the deformation
of the potential $V(Q)$, which is assumed to be included already
in the definition eq.(\ref{pot}).
{}From these considerations we approximate the function $(Z(\tau)-Z(\tau'))$ as
\beqa
Z(\tau)-Z(\tau')=\left\{ \begin{array}{llr} -1 & & |\tau|\geq 2\omega_0^{-1},
|\tau'|\leq \omega_0^{-1} \\
                                             1 & & |\tau|\leq \omega_0^{-1},
|\tau'|\geq 2 \omega_0^{-1}
  \end{array}\right. .\label{Zap}
\eeqa
Using this expression, the time integral in eq.(\ref{DS3}) is
\beqa
\int_{-\infty}^\infty d\tau\int d\tau' \frac{(Z(\tau)-Z(\tau'))^2}
{(\tau-\tau')^2}
&\simeq&
  2\int^{\omega_0^{-1}}_{-\omega_0^{-1}}
d\tau \left[\int^\infty_{2\omega_0^{-1}} d\tau' +
\int_{-\infty}^{-2\omega_0^{-1}}
d\tau'\right] \frac{1}{(\tau-\tau')^2}  \nonumber\\
&=& 4\ln 3=4.394.  \label{timeinteg}
\eeqa
This is very close to the numerical result $\simeq 4.2$ without using
eq.(\ref{Zap}).
For the present problem of tunneling, no cutoff is needed at long time,
in contrast to the case of quantum coherence\cite{LCD,ATF}.
The action now reads
\beq
\Delta  S_{\rm dis}\equiv \eta N\epsilon .\label{DSf}
\eeq
where the
strength of dissipation is
\beqa
\eta&= &\frac{3\ln 3}{16\pi}{(\kfu^2-\kfd^2)^2 a^4}\frac{\lambda}{ a}
\int^{(\kfu+\kfd)\frac{\pi}{2}\lambda}_{(\kfu-\kfd)\frac{\pi}{2}\lambda}
dx\frac{1}{x} \frac{1}{\cosh^2 x}  \nonumber\\
&\equiv&\frac{3\ln 3}{16\pi}{(\kfu^2-\kfd^2)^2 a^4} I(\lambda),
\label{etadef}
\eeqa
where we noted $Q_0=\sqrt{(3/2)}\sqrt\epsilon \lambda$ (eq.(\ref{Qzero})).
The function $I(\lambda)$ is defined by
\beq
I(\lambda)\equiv \frac{\lambda}{ a}
\int^{(\kfu+\kfd)\frac{\pi}{2}\lambda}_{(\kfu-\kfd)\frac{\pi}{2}\lambda}
dx\frac{1}{x}
\frac{1}{\cosh^2 x} .
\eeq
Its asymptotic behaviors are
\beqa
I(\lambda)\simeq
  \left\{ \begin{array}{cr}
      \frac{\lambda}{a}\ln\frac{\kfu+\kfd}{\kfu-\kfd} & \,\,\,
 \lambda(\kfu+\kfd)\ll 1 \\
     \frac{4}{\pi}\frac{1}{(\kfu-\kfd) a}
           e^{-\pi\lambda(\kfu-\kfd)}         &\,\,\, \lambda(\kfu-\kfd)
\gg 1
 \end{array} \right.  \label{Ilambda}
\eeqa
Thus the effect of dissipation is exponentially small for a thick wall
$\lambda(\kfu-\kfd)\gg 1$.
In the case of an bulk iron, for example, the domain wall thickness is about
$\lambda\simeq200 $\AA \ and so the effect of dissipation
from the itinerant electron on MQT would be negligible.
For a thin wall, on the other hand, as realized in {\it e.g.,} SmCo$_{5}$
($\lambda\simeq12$\AA), $\eta$ may be $O(1)$.
To see this, we define dimensionless parameters
\beqa
\ktil&\equiv& \frac{ (\kfu+\kfd)a}{2} \\
\ktil \delta &\equiv& \frac{ (\kfu-\kfd)a}{2} \label{deltadef}.
\eeqa
$\delta$ defined in eq.(\ref{deltadef}) is plotted as a function of
$(U/\epsilon_{\rm F}a^3)$ in Fig. \ref{FIGMF}.
In terms of these parameters and $\lamtil\equiv \lambda/a$,
the strength of dissipation
$\eta$ is written as
\beq
\eta=\frac{3\ln3}{\pi}\ktil^4\lamtil\delta^2
\int^{\pi\ktil\lamtil}_{\pi\ktil\lamtil\delta}\frac{dx}{x}\frac{1}
{\cosh ^2 x}
,  \label{eta}
\eeq
and the approximate expression in the case of weak ferromagnet with
$\pi\ktil\lamtil\delta\ll1$ is
\beq
\eta\simeq \frac{3\ln3}{\pi}\ktil^4\lamtil\delta^2
\left|\ln({\pi\ktil\lamtil\delta})\right|.
\eeq
In Fig. \ref{FIGeta}, $\eta$ evaluated by eq.(\ref{eta}) is plotted as a
function of $\lamtil$ for $\ktil=3$ and $\delta=0.05,0.1$ and $0.2$.
It is seen that $\eta$ can be of the order of 0.1 for $\tilde\lambda\simeq
2$ and $ \delta\simeq 0.05$.
In contrast to the present metallic case, the ohmic dissipation vanishes
in insulators at absolute zero, since in that case,
the major source of dissipation is the magnon, which has a
gap due to the anisotropy.

In the presence of dissipation, the tunneling rate $\Gamma$ is given by
eq.(\ref{rate})  with the action given by eq.(\ref{DSf}).
For a typical case of metal; $ \eta=0.1$, $\Gamma$ is shown in  Fig.
\ref{FIGgamma} for $N=10^4$ and $10^6$ with $h_c=10^{-4}$ .
The value $N=10^4$ corresponds, for instance, to a material with the wall
thichness of $\sim10$\AA\ and the area of $200$\AA$\times 200$\AA.
The rate $\Gamma_0$ without dissipation defined by eq.(\ref{rate0}),
 which corresponds to the case of an insulator are also
plotted by dashed lines.
It is seen that to obtain an observable tunneling rate ($\gsim10^{-5}$Hz) in
metals, smaller value of $\epsilon$ by a factor of about $10^{-2}$ is needed
than in insulators.
%
\section{Dissipation in Multi-band Model}
\label{SECsd}
Our preceeding calculation shows strong dissipative effects for a thin wall
in weak ferromagnet.
In convensional bulk metals, these two conditions might not be easy to be
satisfied simultaneously.
However, in realistic situations, the existence of multi-band ({\it e.g.,}
$s$-$d$ two band model) must be taken into
account, and this is the subject of this section.
It will be shown that dissipation which arises from nonmagnetic conduction
band is effective in the case of a thin magnetic domain wall regardless of
the strength of ferromagnetism.

We consider the simplest case of the multi-band,
the $s$-$d$ model, where the localized magnetic moment is due to $d$
electron and the current is carried by $s$ electron.
Due to the $s$-$d$ mixing, in this case,  the antiferromagnetic exchange
interaction between localized moment $\Mv(\xv)$ of $d$ electrons and the $s$
electron arises\cite{SW}
\beq
H_{sd}=g \int d^3x \Mv(\xv) \cdot (c^{(s)\dagger} {\sigbf} c^{(s)})_\xv .
\label{sd}
\eeq
The Lagrangean of the $s$ electron is given
\beq
L^{(s)}=\sum_{\kv\sigma} c^{(s)\dagger}_{\kv \sigma}(\partial_\tau +
\epsilon_\kv) c_{\kv\sigma}^{(s)}
  +g\int d^3x \Mv(\xv) (c^{(s)\dagger} \sigbf c^{(s)})_\xv .
\eeq
This Lagrangean is of the same form as eq.(\ref{L1}), except that the last
term in eq.(\ref{L1}) is absent here, since the
Coulomb interaction among $s$ electron is neglected.
Due to the localized moment, the $s$ electron in the rotated frame
is polarized as
\beq
(\kfu^{(s)})^2-(\kfd^{(s)})^2 =4\mels gM, \label{plosel}
\eeq
where $\kfs^{(s)}$ and $\mels$ are the fermi momenta and the band mass of the
$s$ electron, respectively, and we have assumed $\Ms\equiv |\Mv|$ is
a constant.
The calculation is carried out in the same way as in \S \ref{SECeff} and
\ref{SECdw}.
The contribution to the local part of the effective action from the
$s$ electron is obtained in the form of eq.(\ref{Sloc}) with $S$ replaced by
$M^{(s)}a^3/2\equiv [(\kfu^{(s)})^3-(\kfd^{(s)})^3]a^3/(12\pi^2)$, and
$JS^2$ by $(\nels/4\mels)
[1-\{(\kfu^{(s)})^5-(\kfd^{(s)})^5\}/(30\pi^2\nels\mels gM)]$,
$\nels$ being the density of $s$ electron.
This contribution renormalizes the magnitude of spin $S$ and exchange $J$,
but this renormalization can be understood as already included in the
values of these quantities, and hence only non-local
term $\Delta S_{\rm dis}^{(s)}$ arising from $s$ electron is considered
below.
Similarly to eq.(\ref{etadef}), the strength of dissipation due to $s$
elctron
is obtained as $(\Delta S_{\rm dis}^{(s)}\equiv \eta^{(s)} N\epsilon)$
\beq
\eta^{(s)}=\frac{3\ln 3}{64\pi}{[(\kfu^{(s)})^2-(\kfd^{(s)})^2]^2 a^4}
\frac{\lambda}{ a}
\int^{(\kfu^{(s)}+\kfd^{(s)})\frac{\pi}{2}\lambda}
_{(\kfu^{(s)}-\kfd^{(s)})\frac{\pi}{2}\lambda}
dx\frac{1}{x} \frac{1}{\cosh^2 x}. \label{etas}
\eeq
This expression is the same form as $\eta$ (eq.(\ref{etadef})), but is
 devided by a factor of four because RPA summation is not needed.
The behavior of $\eta^{(s)}$ is accordingly read from Fig.\ref{FIGeta}.
{}From the discussion after eq.(\ref{eta}), $\eta^{(s)}$  can be $O(0.1)$
if the lower bound of the integration, $(\kfu^{(s)}-\kfd^{(s)})\pi\lambda/2$,
is smaller than unity.
The difference of fermi momenta is written in the case of weak $s$-$d$
coupling  as $(\kfu^{(s)}-\kfd^{(s)})\simeq \kf^{(s)}\alpha$
$(\alpha\equiv (g \Ms/\ef^{(s)})\ll 1)$, where $\ef^{(s)}$ and $\kf^{(s)}$
are the fermi energy and momentum in the absence of $\Ms$.
Therefore, in the $s$-$d$ model, in contrast to the case of single-band
model,
large value of $\eta^{(s)}\simeq0.1$ is possible even in strong ferromagnet,
if the $s$-$d$ coupling is sufficiently small:
$(\lambda\kf^{(s)}\alpha \ll 1)$.
The expression of $\eta^{(s)}$ in this limit is
\beq
\eta^{(s)}\simeq \frac{3\ln 3}{16\pi}(\kf^{(s)} a)^4 \left(\frac{\lambda}{a}
\right) \alpha^2\left|\ln\left(\frac{\pi}{2}\lambda\kf^{(s)}\alpha\right
)\right|.
\eeq
For $\alpha=0.05$, $(\lambda/a)\simeq2$ and $\kf^{(s)} a=3$, for example,
$\eta^{(s)}\simeq 0.02$.
The result indicates that in the multi-band systems,
dissipation due to Stoner excitation will be significant for a thin wall
even in the case of strong ferromagnet.

\section{Dissipation due to Eddy Current}
\label{SECeddy}
Besides the direct coupling to the electron spin current, the moving wall in
metals also interacts with the charge current (eddy current) via induced
 electric field governed by Faraday's law as noted by Chudnovsky
{\it et al.}\cite{CIS}.
The dissipation due to this eddy current is ohmic, but this effect turns
out to be small for a case of meso- or microscopic wall, as discussed below.

The electric field induced due to the change of the magnetization is
calculated
 from the Maxwell equation
\beq
\nabla\times{\bf E}=-\mu_0\dot{\bf M},
\eeq
where ${\bf M}\equiv (2\hbar\gamma S/a^3)\nv$ is the magnetization
vector.
For the case of a moving domain wall, where $\nv$ is given by
eq.(\ref{DW}), only the $y$-component $ E_y$ is important, which is obtained
as
\beqa
E_y(\xv,\tau)\simeq\mu_0\left(\frac{2\hbar\gamma S}{a^3}\right)
\dot Q(\tau) \times \left\{ \begin{array}{lr}
  \tanh\frac{x-Q}{\lambda} & |x-Q| \lsim L  \\
  \frac{1}{2\pi}\frac{L^2}{(x-Q)^2} {\rm sgn}(x-Q) & |x-Q| \gg L \end{array}
  \right., \label{electric}
\eeqa
where $L$ is the linear dimension of the crosssection of the wall
($A_w=L^2$).
For $L\gg \lambda$, the electric field is almost constant over the distance
$|x-Q|\lsim L$ and decays for a larger distance.

The electromagnetic coupling to electrons is
\beq
H_{\rm EM}=e\int d^3x {\bf A}\cdot{\bf j},
\eeq
where the electron current is given by
\beq
{\bf j}(\xv,\tau)=-\frac{i}{2m}(c^\dagger\nabla c-\nabla c^\dagger c),
\eeq
and the vector potential is related to the electric field by
${\bf E}=-\dot{\bf A}$.
Since ${\bf E}$ and hence ${\bf A}$ depend on $Q$, this coupling
$H_{\rm EM}$
gives rise to dissipation effects on the motion of $Q$.

Treating $H_{\rm EM}$ perturbatively to the second order and integrating
out the electron, we obtain the following term in the effective action for
$Q$;
\beq
\Delta S_{\rm ch}=-\frac{1}{2}\int d\tau\int d\tau' \frac{1}{V}\sum_\qv
 A_\mu(\qv,\tau)A_\nu(-\qv,\tau')<j_\mu(\qv,\tau)j_\nu(-\qv,\tau')>,
\eeq
where $A_\mu(\qv,\tau)$ and $j_\mu(\qv,\tau)$ are the Fourier transforms
\beqa
A_\mu(\qv,\tau)&=&\int d^3 x e^{-i\qv\xv}A_\mu(\xv,\tau) \nonumber\\
j_\mu(\qv,\tau)&=&\frac{1}{\sqrt V}\sum_{\kv\sigma}\frac{1}{m}
\left(k_\mu+\frac{q_\mu}{2}\right) c^\dagger_{\kv+\qv,\sigma}
c_{\kv,\sigma}(\tau).
\eeqa
{}From the behavior of $ E_y(\xv,\tau)$, it is easily seen that $A_y(\qv,\tau)$
is well approximated as
\beqa
A_y(\qv,\tau)\simeq \left\{ \begin{array}{cc}
 \mu_0\left( \frac{2\hbar \gamma S}{a^3}\right)\frac{1}{iq_x}e^{iq_x Q}
A_w L \times \delta_{q_y,0}\delta_{q_z,0} & (q_x\lsim L^{-1}) \\
0 & (q_x\gg L^{-1})  \end{array}\right. .
\eeqa

We consider the disordered case where $L$ is much larger than the mean
free path of electron $\ell$;
\beq
L\gg \ell,
\eeq
since this is of most interest in experimental situations.
Because of this condition, the correlation function
$<j_\mu(\qv,\tau)j_\nu(-\qv,\tau')>$ can be approximated by its value at
$q=0$.
In this case, the ohmic contribution is written in terms of the conductivity
$\sigma$ as
\beqa
<j_\mu(\qv,\tau)j_\nu(-\qv,\tau')>&\simeq& \frac{\sigma}{\pi}\int_0^\infty
\omega d\omega
e^{-\omega|\tau-\tau'|}  \nonumber\\
&&=\frac{\sigma}{\pi(\tau-\tau')^2}.
\eeqa
The action is thus written as
\beq
\Delta S_{\rm ch}\simeq \frac{1}{4\pi}\left[\mu_0\left(\frac{2\hbar\gamma S}
{a^3}\right)\right]^2 \sigma A_w L\times \int d\tau\int d\tau'
\frac{(Q(\tau)-Q(\tau'))^2}{(\tau-\tau')^2},
\eeq
where we have used $(Q(\tau)-Q(\tau'))/L\ll 1$
and subtracted an irrelevant constant.
Therefore the strength of ohmic dissipation $\eta^{\rm (ch)}$ due to the
eddy current, defined by $\Delta S_{\rm ch}=\eta^{\rm (ch)}N\epsilon$
($N=A_w\lambda/a^3$ and $\epsilon$ arises
from $Q_0\simeq \sqrt\epsilon \lambda$), is
\beq
\eta^{\rm (ch)}\simeq\frac{1}{4\pi}\left[\mu_0\left(
\frac{2\hbar\gamma S}{a^3}
\right)\right]^2 \frac{a^5}{\hbar} \sigma \left(\frac{\lambda}{a}\right)
\left(\frac{L}{a}\right).
\label{etach}
\eeq
For a typical metal, $\sigma\simeq 10^8$[$\Omega^{-1}$m$^{-1}$],
it is  $\eta^{\rm (ch)}\simeq 10^{-7}\times ({\lambda}/a)({L}/{a})$
if we choose $S\simeq1$ and $a=3\times 10^{-10}$[m].
Hence, as far as $(\lambda/a)\lsim 10^2$, which is commonly satisfied,
dissipation due to eddy current
is small in mesoscopic wall we are interested in; $(L/a)\lsim 10^4$.

Let us remark on a useful way of estimating $\eta^{\rm (ch)}$.
By the definition of $\eta^{\rm (ch)}(\equiv \Delta S_{\rm ch}/(N\epsilon))$,
we can write
\beq
\Delta S_{\rm ch}= N\frac{\eta^{\rm (ch)}}{\lambda^2} \int d\tau\int d\tau'
\frac{(Q(\tau)-Q(\tau'))^2}{(\tau-\tau')^2}.
\eeq
Noting that the integrand is regarded as rate of energy dissipation due to
the Joule heat, we have the relation
\beq
\frac{N}{V}\frac{\eta^{\rm (ch)}}{\lambda^2}\dot Q^2 \simeq \sigma {\bf E}^2,
\eeq
where $V=A_w L$.
By use of eq.(\ref{electric}), we obtain the expression eq.(\ref{etach})
corrent up to a numerical factor.

\section{Discussions}
\label{SECdisc}
We have studied the case of a domain wall where the plane of the wall is
parallel to the easy ($yz$-)plane.
It is also possible to consider based on the action eq.(\ref{Smag})
another configuration where spin direction changes in the easy ($z$-)
direction of spin as shown in Fig. \ref{FIGDW}(b), as long as the anisotropy
energy $K$ is much larger than the demagnetization energy $K_d$.
For the MQT of this wall, the results obtained above do not change.

We have neglected the effect of magnetic field on electronic states.
This is justified as long as $UM\gg\gamma H$.
In experimental situations with the magnetic field of $\lesssim1$T and
$U\simeq 10$eV,
this condition reduces to $M\gtrsim 10^{-4}$ in unit of the Bohr magneton,
which is easy to satisfy.
However, in order to discuss the case of very small $M$, the fluctuation
of the magnitude $M_{\bf x}$ around the mean field value must also be
included.

The contributions of higher order in $H_{\rm int}$ are smaller than that of
the second order we have calculated; for the potential
renormalization by the order of $(k_{{\rm F}\uparrow}\lambda)^{-2}$ and
for the dissipative effect
by $(k_{{\rm F}\uparrow}\lambda)^{-2}$ or $\epsilon$.

In Eq. (\ref{eta}) we have taken account of only the ohmic dissipation.
The super-ohmic contributions, which are of higher orders of
$(\omega_0/\epsilon_{\rm F})$ in Eq. (\ref{Im}), are
smaller than the ohmic one by a factor of
$(\omega_0/\epsilon_{\rm F})^2 \ll 1$ and hence are negligible.
On the other hand, a contribution from the magnon pole leads to the
dissipation with a gap $\Delta_0$ due to anisotropy, as
discussed in Appendix\ref{SECswpole}.
In the present problem of tunneling, the effect needs to be evaluated
quantitatively. This is because, in contrast to the case of quantum coherence
problem\cite{LCD,ATF}, the ohmic dissipative action is not divergent
at long time, and hence the contribution from the ohmic
dissipation is not qualitatively distinct than those from the
super ohmic one and dissipation processes with an excitation gap.
The strength of dissipation $\eta^{\rm (pole)}$, which is
defined by the value of the action devided by $(N\epsilon)$, is obtained as
(eq.(\ref{etasw}))
\beq
\eta^{\rm (pole)} =\frac{3\pi}{40}(\Ms a^3)e^{-\frac{\Delta_0}{\omega_0}}.
\eeq
Since experiments are usually carried out in highly anisotropic materials
with
 $\Delta_0/\omega_0\simeq 10$, this contribution is very small
compared to the ohmic dissipation for the case of a thin wall.

Our result shows a distinct difference between MQT of thin
walls in metallic and insulating magnets.
Unfortunately the experiments carried out so far appear not yet be able to
test the theoretical prediction of the effect of dissipation due to
itinerant electrons.
The experimental result on small ferromagnetic particles of
Tb$_{0.5}$Ce$_{0.5}$Fe$_2$ suggests the motion of a domain wall via MQT
below $T_{co} \simeq 0.6$K\cite{PSB}.
In this experiment, the width of the domain wall is about $30$\AA\ and
according to our result, $\eta\propto \exp[-\pi\lambda(k_{{\rm F}\uparrow}
-k_{{\rm F}\downarrow})]$,
the dissipation from electron spin current is negligible for such a thick
wall.
This may be the reason why the result of the crossover temperature
$T_{\rm co}\sim0.6$K is roughly in agreement  with the theory\cite{Sta}
  without dissipation
 (eq.(\ref{Tc0}) with $h_c \sim 4\times 10^{-4}$ and $\epsilon \sim0.1$).
On the other hand, the domain wall in SmCo$_5$ is
very thin $\lambda\simeq12$\AA, and our
result suggests strong effect of dissipation due to Stoner excitation,
which will be interesting to observe.
Experiments on bulk crystal of
SmCo$_{3.5}$Cu$_{1.5}$ with very thin walls (a few times $a$) have
been performed\cite{UB}, although quantitative comparison
is not easy since many walls will participate in the relaxation processes of
the magnetization in these experiments.
Even in the case of thick walls, the dissipative effect becomes large in
weak ferromagnets, where the experiments, however, will not be easy
because of small value of saturation magnetization $M$.

MQT in disordered magnets has a new  possiblity of observing a
significant effect of sub-ohmic dissipation.
In fact, as disorder is increased in a metallic magnet, the Anderson
transition
into an insulator will occur, and it was shown recently that near the
transition the  dissipation  due to the
conduction electron becomes sub-ohmic\cite{ATF}.
Disordered magnets may also be suitable for study of MQT because
the effect of eddy current becomes less important
for larger resistivity as seen from Eq. (\ref{etach}).

\section{Conclusion}
\label{SECconc}
We have studied the macroscopic quantum tunneling of a
domain wall in a metallic
ferromagnet based on the Hubbard model.
Integrating out the electron degrees of freedom by use of the locally rotated
frame, we obtained the effective action of the magnetization.
The term local in time (instantaneous) in the effective action has the same
form as that of a ferromagnetic Heisenberg model, {\it i.e.,} there is no
formal difference between metals and insulators.
On the other hand, the non-local (retarded) part, which describes the
dissipative effect on magnetization, is crucially different from the
insulating case.
Because of the gapless Stoner excitation in the itinerant electron, the ohmic
dissipation exists even at zero temperature.
The effect is negligible for a thick domain wall where experiments so far
have
been carried out.
On the other hand, important effects of the ohmic dissipation are expected
in
thin domain walls with thickness $\lambda$ comparable
to the inverse of the difference of the fermi momenta $(\kfu-\kfd)^{-1}$.
In the strong ferromagnets, the ohmic dissipation arising from the magnetic
band is weakend. In this case the existence of multi-band needs to be taken
into account since the dissipation due to non-magnetic band is significant
even in strong ferromagnets.
We believe the observation of the dissipation is within the present
experimental attainability.

In our analysis we have not dealt with the very weak ferromagnetizm.
In this case, the magnitude of the magnetization, in addition to its
direction, might change in the domain wall and further studies are needed.
Our formulation can be easily extended to the case of MQT of the
magnetization vector\cite{CG} in a single domain metallic ferromagnet.
This will be investigated in the forthcoming paper.

\acknowledgements
The authors are grateful to M. Hayashi, H. Kohno and H. Yoshioka for
valuable discussions.
G. T. also thanks K. Nosaka for her assistance in collecting articles.
This work is financially supported by Ministry-Industry Joint Research
program "Mesoscopic Electronics" and by Grant-in-Aid for Scientific Research
on Priority Area, "Electron Wave Interference Effects in Mesoscopic
 Structure"
(04221101) and by Monbusho International Scientific Research Program:Joint
Research "Theoretical Studies on Strongly Correlated Electron Systems"
(05044037) from the Ministry of Education, Science and Culture of Japan.



\appendix
\section{The Locally Rotated Frame for Electron}
\label{AProt}
We will explain the locally rotated frame we have used to separate the fast
varying degrees of freedom from the slowly varing one and
in deriving the effective action for the latter.
For simplicity we consider the conduction electrons coupled to a local
magnetization vector $\Mv(\xv)$
\beq
L=\sum_{\kv\sigma} c^\dagger_{\kv \sigma}(\partial_\tau + \epsilon_\kv)
c_{\kv\sigma}  -g\sum_{\xv}\Mv_\xv (c^\dagger {\bf \sigbf} c)_\xv .
\eeq
The configuration of the magnetization $\Mv(\xv)\equiv \Ms \nv(\xv,\tau)$ is
a domain wall given by eq.(\ref{DW}).

We will first demonstrate that a naive perturbation in terms of $g$ is not
appropriate in calculating the effective action.
The dissipative term in the second order of $g$ is given by
\beq
\Delta {S}_{\rm dis}^{\rm (pert)}=-\frac{g^2}{2}\int d\tau \int d\tau'
\int d^3x \int d^3 x'\sum_{ij} M^i(\xv,\tau)M^j(\xv',\tau')
\frac{1}{V}\sum_q e^{-i\qv(\xv-\xv')}\frac{1}{\beta}\sum_{\omega_\ell}
e^{i\omega_\ell(\tau-\tau')}\chi^{ij}_{q\ell},
\eeq
where the spin correlation function,
\beq
\chi^{ij}_{q\ell}\equiv \frac{1}{V}\sum_{\kv\kv'}\frac{1}{\beta}
\sum_{\omega_n,\omega_{n'}}<(c^\dagger_{\kv+\qv,n+\ell} \sigma^i c_{\kv,n})
(c^\dagger_{\kv'-\qv,n'-\ell} \sigma^j c_{\kv',n'}) > ,
\eeq
is calculated using the unpolarised electron; $\epsilon_{\kv}=
\kv^2/(2m)-\epsilon_{\rm F}$.
Picking up the $\omega$-linear terms of Im$\chi^{ij}_{q\ell}
(i\omega_\ell\equiv \omega+i0)$ that governs the low energy behavior
(similary to the calculation in \S\ref{SECeff}),
$\Delta {S}_{\rm dis}^{\rm (pert)}$ is reduced to
\beqa
\lefteqn{ \Delta {S}_{\rm dis}^{\rm (pert)}=N\frac{\kf^4 \lambda a^3}{16}
\int d\tau\int d\tau' \frac{1}{(\tau-\tau')^2}  }\nonumber\\
&&\times\int\frac{dq_1}{2\pi}\frac{ \sin^2\frac{q_1}{2}(Q(\tau)-Q(\tau'))}
{|q_1|}
\left( \frac{1}{\cosh^2 \frac{\pi}{2}\lambda q_1}   + \frac{1}{\sinh^2
\frac{\pi}{2}\lambda q_1} \right) \theta(2\kf-|q_1|) .\label{DSpert}
\eeqa
Here  factors $1/{\cosh ({\pi}\lambda q_1/2)}$ and $1/{\sinh ({\pi}\lambda
q_1/2)}$ are the form factors of $M^y_\xv$ and $M^z_\xv$, respectively;
\beqa
\int d^3\xv e^{-i\qv\xv}M_\xv^y &=&
 \pi \Ms \frac{A_{\rm w}\lambda}{a^3}
e^{-iq_1 Q(\tau)}\frac{1}{\cosh \frac{\pi}{2}\lambda q_1}
 \delta_{q_20}\delta_{q_3 0}  \\
\int d^3\xv e^{-i\qv\xv}M_\xv^z &=&
 -i\pi \Ms \frac{A_{\rm w}\lambda}{a^3}
e^{-iq_1 Q(\tau)}\frac{1}{\sinh \frac{\pi}{2}\lambda q_1}
 \delta_{q_20}\delta_{q_3 0}.
\eeqa
It is seen that the contribution from $M^z$ diverges logarithmically  at
$q_1\rightarrow0$.
This implies that the perturbation with respect to $g$ is not valid.

On the other hand, if $\Delta  S_{\rm dis}$ is calculated in the rotated
frame
(eq.(\ref{DS3})), where the contribution in the first order in $g$ is taken
into account in the unperturbed state, it reads
\beqa
\Delta  S_{\rm dis}^{\rm (rot)} &\simeq& \frac{1}{4}\times N
\frac{m^2}{2\pi}(gM)^2\frac{a^3Q_0^2}{\lambda}
 \int d\tau\int d\tau' \frac{(Z(\tau)-Z(\tau'))^2}{(\tau-\tau')^2}
\nonumber\\
&&
\times \int^{(\kfu+\kfd)\frac{\pi}{2}\lambda}_{(\kfu-\kfd)\frac{\pi}{2}
\lambda}
\frac{dx}{x} \frac{1}{\cosh^2 x} ,
\eeqa
where we have used $\kfu^2-\kfd^2 =4mgM$ $(M\equiv |\Mv|$) and a factor of
$1/4$ is due to the absence of RPA summation.
According to eq.(\ref{nm}), the fermi momentums are written at
$g/\epsilon_{\rm F}\rightarrow 0$ as
\beq
\kfs\simeq \kf (1\pm \half\frac{g}{\epsilon_{\rm F}}M),
\eeq
where $\ef$ and $\kf$ are the fermi energy and momentum at $g=0$,
respectively, and so
\beqa
\Delta  S_{\rm dis}^{\rm (rot)} &\simeq& N \frac{1}{32\pi}\left(\frac{g}
{\epsilon_{\rm F}}M\right)^2 \kf^4\frac{a^3}{\lambda}Q_0^2
 \int d\tau\int d\tau' \frac{(Z(\tau)-Z(\tau'))^2}{(\tau-\tau')^2}
\times \int^{\pi\lambda\kf}_{\frac{\pi}{2}\lambda \kf
\frac{g}{\epsilon_{\rm F}}M}\frac{dx}{x} \frac{1}{\cosh^2 x} \nonumber\\
&\simeq&  N \frac{1}{32\pi} \kf^4\frac{a^3}{\lambda}Q_0^2
 \int d\tau\int d\tau' \frac{(Z(\tau)-Z(\tau'))^2}{(\tau-\tau')^2}
\times\left(\frac{g}{\epsilon_{\rm F}}M\right)^2\left|\ln\left(
 \frac{g}{\epsilon_{\rm F}}M\right) \right|     ,
\eeqa
The logarithmic correction  arises from the lower bound of momentum
integration. Hence the difference of the fermi momenta, $(\kfu-\kfd)\simeq
(\kf gM/\ef)$, works as a small momentum cutoff.
Thus the expansion parameter at small coupling is not
$(gM/\epsilon_{\rm F})^2$ but $(gM/\epsilon_{\rm F})^2
|\ln(gM/\epsilon_{\rm F})|$.
This is the reason why naive perturbation breaks down.
%
%
\section{A Case of a Ferromagnet with Uniaxial Anisotropy}
\label{demag}
We have considered in \S\ref{SECdw1} the case of a ferromagnet with
anisotropy energies in $z$- and $x$-direction.
In the case of uniaxial anisotropy, the same action, eq.(\ref{Smag}),
can also be a starting point, if the effect of demagnetization
is taken into account, as we shall discuss here.
The Hamiltonian with uniaxial anisotropy is
\beq
H_0=\int d^3x \left[\frac{JS^2}{2}\left((\nabla\theta)^2+\sin^2
(\nabla\phi)^2
\right)-\frac{KS^2}{2}\cos^2\theta \right].
\eeq
We will deal with only a classical domain wall solution of this system that
is planar, namely, the spin configuration changes only in one
direction, say, in $x$-direction, and hence it is uniform in the $y$-$z$
plane.
The spatial variation of the magnetization vector,
${\bf M}\equiv 2\hbar\gamma{\bf S}/a^3$[A/m]
($\gamma=e/(2m)$ being the gyromagnetic ratio),
gives rise to the magnetic pole
and hence induces the demagnetization field ${\bf H}_{\rm d}$ via
 Maxwell equation $\nabla \cdot {\bf H}_{\rm d}=-\nabla \cdot {\bf M}$.
In the case of the planar domain wall we are considering, the demagnetization
field points in the $x$-direcxtion and is expressed by the $x$-component of
the magnetization as ${\bf H}_{\rm d}=(-M_x, 0,0)$.
The magnetostatic energy due to the demagnetization field
 is thus expressed in the form of an anisotropy energy for the spin
in the $x$-direction;
\beq
E_{\rm d}\equiv \frac{\mu_0}{2}\int d^3x \; {\bf H}_{\rm d}^2
 =\frac{\mu_0}{2} \frac{(2\hbar\gamma)^2}{a^6} S^2 \int d^3x
 \sin^2\theta \cos^2 \phi
\eeq
where $\mu_0$ is the magnetic permeability of free space.

Hence the effective Hamiltonian describing the ferromagnet with
 uniaxial anisotropy is equivallent
to the action considered in \S\ref{SECdw1} (eq.({\ref{Smag})),
if the effect of the demagnetization is taken into account, but with
the transverse anisotropy replaced by the demagnetizaion energy;
$K_{\perp}\rightarrow \mu_0(\hbar\gamma)^2/a^6$.

For the case of weak transverse anisotropy energy compared to the
demagnetization energy ($K_{\perp}\ll\mu_0(\hbar\gamma)^2/a^6$),
as realized in most conventional bulk metals, one should also read
$K_{\perp}$ as $\mu_0(\hbar\gamma)^2/a^6$.

\section{The Dissipation from Spin Wave Excitation}
\label{SECswpole}
We have calculated the ohmic dissipation  arising from
the Stoner excitation as given by eq.(\ref{Sdis}).
There is also dissipation due to the spin wave (magnon),
which is present also in the insulating ferromagnet,
but this is not ohmic at $T=0$ because of the spin wave gap due to anisotropy.
This contribution is estimated below.

For small momentum and energy transfer, $q\ll (m/\kf)U\Ms$ and
$\oml\ll U\Ms$,
$C_{+-}^{ij}$ in the current correlation function $<\Jv_+\Jv_->$ of
eq.(\ref{JJpm}) is proportional to
\beq
\frac{2U}{1-2U\chi^0_{+-}(q,\ell)}\simeq -\Ms(2U)^2 \frac{1}{i\oml-\omega_q},
\eeq
where $\omega_q\equiv\Delta_0+J\Ms a^6q^2/2$ with the magnon gap given by
$\Delta_0\equiv Ka^6{\Ms}/{2}$.
The behavior of $C^i_{+-}(q,\ell)$  at small $q$ and $\omega_\ell$
is obtained from eq.(\ref{Imc}) as
\beq
C^i_{+-}(q,\ell)\simeq -q_i\frac{J\Ms a^6}{4U}+O(q^3,\omega_\ell).
\eeq
The imaginary part of $<\Jv_+\Jv_->$ arising from the spin wave pole is thus
\beqa
{\rm Im}<J^i_+J^i_->|^{\rm (sw)}_{\omega+i0} &\simeq& -\Ms (J\Ms a^6)^2
\frac{q_i^2}{4}  {\rm Im}\frac{1}{\oml-\omega_q+i0} \nonumber\\
&=& \pi \Ms (J\Ms a^6)^2\frac{q_i^2}{4} \delta(\oml-\omega_q).
\eeqa
By use of eqs.(\ref{DSexamp}) (\ref{AC}), the spin wave contribution to the
action $\Delta S_{\rm dis}$ is
\beq
\Delta S_{\rm dis}^{\rm (sw)}=\frac{1}{32}\Ms^3 J^2 a^{12}
\int\! d\tau \! \int \! d\tau'  \frac{1}{V}\sum_{\qv} q_x^2
|\Theta_\qv^1(\tau)-\Theta_\qv^1(\tau')|^2
 e^{-(\Delta_0+J{\Ms} a^6q^2/2)|\tau-\tau'|}.
\eeq
Because of anisotropy gap, the $(\tau-\tau')$-dependence of the correlation
function is $\exp(-\Delta_0|\tau-\tau'|)$ at long time instead of
$(\tau-\tau')^{-2}$ for the ohmic case.
Using the bounce solution (\ref{thetbounce}) for $\Theta_\qv^1(\tau)$,
the action reduces to
\beqa
\Delta S_{\rm dis}^{\rm (sw)} & \simeq &
N\frac{\pi^2}{8}(Ja)^2(\Ms a^3)^3\frac{a^4}{\lambda }
\int d\tau\int d\tau' e^{-\Delta_0|\tau-\tau'|} \nonumber\\
&&\times\int\frac{dq}{2\pi} q^2\sin^2\frac{q}{2}(Q(\tau)-Q(\tau'))
\frac{1}{\cosh^2 \frac{\pi}{2}\lambda q} \nonumber\\
&\equiv& \eta^{\rm (sw)} N\epsilon ,\label{DSsw}
\eeqa
where $\eta^{\rm (sw)}$ is given by
\beqa
\lefteqn{ \eta^{\rm (sw)}  \simeq
\frac{3\pi}{160}\left(\frac{Ja}{\Delta_0}\right)^2 (\Ms a^3)^3
\left( \frac{a}{\lambda}\right)^4
 e^{-\frac{\Delta_0}{\omega_0}}       }\nonumber\\
&& =\frac{3\pi}{40}(\Ms a^3)e^{-\frac{\Delta_0}{\omega_0}}.  \label{etasw}
\eeqa
In deriving this expression,  we have used $J=\lambda^2 K$, and the fact that
the $q$-integration in eq.(\ref{DSsw}) is dominated by $q\simeq\lambda^{-1}$.
The time integration has been estimated by introducing a short time cutoff of
$\omega_0^{-1}$ similarly to eq.(\ref{timeinteg}) as
\beq
\int_{-\infty}^\infty d\tau\int d\tau'
{(Z(\tau)-Z(\tau'))^2}e^{-\Delta_0 |\tau-\tau'|}
\simeq \frac{4}{\Delta_0^2}e^{-(\Delta_0/\omega_0)},
\eeq
where $Z(\tau)=Q(\tau)/Q_0$.
The experiments are carried out on materials with strong anisotropy
$\Delta_0\sim 10$K\cite{PSB}, while $\omega_0\sim 1$K, and according to
eq.(\ref{etasw}),
the spin wave contribution to dissipation is negligibly small.

\begin{figure}

\caption{
  Polar coordinates $(\theta,\phi)$ that represent the direction $\nv$
 of the magnetization vector $\Mv$.
  \label{FIGpolar}}

\caption{
  The diagramatic expression of the spin current correlation functions
$<\Jv_+\Jv_->$ (a) and $<\Jv_z\Jv_z>$ (b) in RPA.
  \label{FIGRPA} }

\caption{
  Mean field solutions of quantities
$\tilde M\equiv (\Ms a^3)$, $\tilde \kfs\equiv (\kfs a)$,
$\tilde J\equiv (J\Ms^2 a^7/\epsilon_{\rm F})$,
$\tilde m_{M}^2\equiv (m_M^2 /\ef a^3)$ and $\delta\equiv(\kfu-\kfd)/
(\kfu+\kfd)$ as a function of $\tilde U\equiv (U/\epsilon_{\rm F}a^3)$.
The ferromagnetism appears for $\tilde U\geq(2/3) $ and the complete
ferromagnetism is realized for $\tilde U\geq1$, where $\kfd$ vanishes.
  \label{FIGMF}}

\caption{   The contours of integrations in the complex $\omega$-plane.
  \label{FIGcont} }

\caption{ A  configuration of magnetization of a planar domain wall (a)
where the spin configuration spatially varies in $x$-direction, which is
 perpendicular to the easy plane of the spin.
 A planer wall with the easy axis perpendicular to the wall is shown in
(b), which is possible if the anisotropy energy along the easy axis
is much larger than the demagnetization energy.
  \label{FIGDW} }

\caption{
  The potential $V(Q)$ for the wall coordinate $Q$ produced by the pinning
center
 and the magnetic  field. The width of the barrier and the attempt frequency
at the local minimum are denoted by $Q_0$ and $\omega_0$, respectively.
The trajectory $Q(\tau)$ of the bounce solution is shown schematically.
  \label{FIGpot} }

\caption{
  The strength of the dissipation $\eta$ given by eq.(\protect\ref{eta})
as a function of the width of the wall $(\lambda/a)$,
for $\delta\equiv (\kfu-\kfd)a/(2\tilde k_0)= 0.05, 0.1$ and $0.2$ with
$\tilde k_0 \equiv(\kfu+\kfd)a/2 =3.0$.
  \label{FIGeta} }

\caption{
  The tunneling rate $\Gamma$ (solid line) for a typical case of an metal
$(\eta=0.1)$ with total number of spins in the wall
$N=10^4$ and $10^6$, respectively. The dashed lines are the rate $\Gamma_0$
in the absence of dissipation.
The parameter is taken as $h_c=10^{-4}$.
  \label{FIGgamma} }

\end{figure}
\end{document}